\documentstyle[12pt,aasms4]{article}

\def\lsim{\lower 0.5ex\hbox{${}\buildrel<\over\sim{}$}}
\def\gsim{\lower 0.5ex\hbox{${}\buildrel>\over\sim{}$}}
\def\VVm{\langle V/V_{max}\rangle}
\def\cost{\langle$cos$\Theta \rangle}
\def\coss{\langle$cos$^2\Theta \rangle}
\def\cosq{\langle$cos$^4\Theta \rangle}
\def\sinb{\langle$sin$^2 b \rangle} 
\def\phivmax{\phi_{\scriptscriptstyle V}^{max}}
\def\vgal{\overrightarrow{V_\Omega}}
\def\vrec{\overrightarrow{V_r}}
\def\taugrb{\tau_{\scriptscriptstyle GRB}}

\def\td{t_{\scriptscriptstyle D}}

\begin{document}

\title{The Halo Beaming Model for Gamma-Ray Bursts}

\author{R.C. Duncan\altaffilmark{1} \&  Hui Li\altaffilmark{2}}

\altaffiltext{1}{Dept.~of Astronomy and McDonald Observatory, 
Univ.~of Texas, Austin, TX 78712; duncan@astro.as.utexas.edu}
\altaffiltext{2}{MS D436, Los Alamos National Laboratory, 
Los Alamos, NM 87545; hli@lanl.gov}

\begin{abstract}
We consider a model for gamma-ray bursts (GRBs)
from high-velocity neutron stars in the galactic halo.
In this model, bursters are born in the galactic disk with large recoil 
velocities ${\bf V_r}$, and  GRBs are beamed to within emission cones 
of half-angle $\phi_b$ centered on $\bf V_r$.  We describe scenarios for 
magnetically-channeled GRBs that have such beaming characteristics.
We then make detailed comparisons of this halo beaming model (HBM) to 
data from the 3rd BATSE Catalog and from the Pioneer Venus Orbiter 
experiment, for both GRB intensity and angular 
position distributions.  Acceptable fits to observations of over 
1000 bursts are obtained for $\phi_b =15^{\circ} - 30^{\circ}$ 
and for a BATSE sampling depth of $D\sim 180$ kpc, which
corresponds to a peak burst luminosity of $\sim 10^{40}$ ergs s$^{-1}$. 
Present data favor a truly isotropic (cosmological) model over the 
HBM, but not by a statistically compelling margin ($\lsim 2 \sigma$).  
 
The HBM makes the distinctive prediction that the galactocentric
quadrupole moment $\langle \hbox{cos}^2 \Theta \rangle - 1/3$ 
for bright, nearby GRBs is large, even though the 
dipole moment $\langle \hbox{cos} \Theta \rangle$ remains near zero. 
Bursters born in nearby external galaxies, such as M31, are almost entirely
undetectable in the HBM because of misdirected beaming.
We analyze several refinements of the basic HBM: gamma-ray intensities
that vary with angle from the beam axis; 
non-standard-candle GRB luminosity functions; and models including
a subset of bursters that do not escape from the galaxy. 
We also discuss the energy budgets for the bursters, the origins of their
recoils, and the physics of burst beaming and alignment. 
One possible physical model is based on the magnetar model of 
soft gamma repeaters (SGRs).  Empirical bounds on the rate of 
formation and peculiar velocities of SGRs imply that there exist 
$\sim 10^4$ to $\sim 10^7$ aged SGRs in the galactic halo
within a distance of 100 kpc.  

The HBM gives an acceptable fit to observations
only if it satisfies some special conditions ($\phi_b \approx20^\circ$, 
uniform bursting rate) which are possible, but for which there are no clear and 
compelling theoretical justifications.  
The cosmological burst hypothesis is more generic and thus more 
attractive in this sense.
\end{abstract}


\section{Introduction}

The Burst and Transient Source Experiment (BATSE) on the NASA Compton
Gamma-Ray Observatory has revealed a nearly isotropic but inhomogeneous 
sky distribution of gamma-ray bursts (Meegan et al.~1992; \cite{mee96}). 
An additional major constraint on GRB theories, 
revealed by the Pioneer Venus Orbiter (PVO) experiment, is that the 
intensity distribution of  bright bursts is consistent with a locally-uniform 
density of sources in Euclidean space (\cite{fen93a}). These facts are 
simply and generically accounted for if the faintest observed GRBs come 
from cosmological distances (e.g., Paczy\'nski 1995).  An alternative 
possibility, that we will focus on here, is that the bursts come from 
the extended halo or corona of our galaxy.  Galactic halo models for GRBs 
were first considered and tested against data by Fishman et al.~(1978) 
and Jennings \& White (1980).  Reasons for favoring halo models 
have been summarized by Lamb (1995). 

In galactic halo GRB models, 
it is often supposed that the bursters are neutron stars, 
since such compact stars are conjectured to 
be capable of producing intense fluxes of hard, non-thermal 
photons; while their small size could drive burst variability 
on submillisecond time scales.
Other reasons for favoring neutron stars include observations of
spectral lines, which are at present controversial, and a
possible connection of classic GRBs with the March 5, 
1979 gamma ray burster (e.g., \cite{dt92}, hereafter DT92) 
which was localized to an angular position lying within a young
supernova remnant (\cite{cetal82}; \cite{rkl94}).
The displacement of this burster from the center of the supernova
remnant indicates that it acquired a velocity $\sim 1000$ km s$^{-1}$
at birth (DT92). Neutron stars with 
such velocities will escape the galactic disk, and move into the halo.
Recent analysis of the 1979 March 5th event lends support to 
its association with classical GRBs (\cite{fkl96}, hereafter FKL).
In addition, many high-velocity radio pulsars ($V > 500$ km s$^{-1}$) 
have been observed (\cite{lyl94}; \cite{fra96}). 

For these reasons, we will focus on theories of GRBs 
from high-velocity neutron stars (HVNSs) 
born in the galactic disk, as first suggested by Shklovskii \&
Mitrofanov (1985).  In particular, we will consider the 
``halo beaming model'' (HBM) proposed earlier 
(\cite{dlt93}, hereafter DLT; \cite{ldt94}, LDT; Li \& Duncan, 1996a,
1996b; Bulik \& Lamb 1996). 
Alternative models for GRBs from HVNSs in the galactic halo
invoke a delayed onset of bursting activity at a time $\sim 10^7$ years 
after birth in the disk (\cite{liderm}, hereafter LD92), which has been 
applied in several different physical contexts 
(\cite{cl94}; \cite{lbc96}; \cite{wh96});
and weakly-bound bursters orbiting in
a nonspherical galactic potential (\cite{prr95}, hereafter PRR). 
Models for bursters {\it born} in the galactic
halo (\cite{es92}; \cite{h92}; \cite{sw93}; \cite{ws94}),
or the Magellanic Clouds (\cite{fp93}),
are beyond the scope of this paper.

In \S \ref{model}, we will describe our basic model for the galactic 
population of bursters, and briefly review its physical motivations. 
In \S \ref{sec-montecarlo} we present our basic Monte Carlo model 
results. We discuss how 
these results arise, and what they could imply for halo GRB theories. 
We also explore the model's sensitivity to different beaming angles.
In \S \ref{sec-hbm-bat-pvo}
we make detailed quantitative comparisons of the model with 
BATSE and PVO data.  In particular, we study moments of the angular position 
distribution in bright subsets of observed GRBs, which is potentially
a sensitive model discriminant. 
In \S \ref{sec-eng-rep} we estimate the GRB repetition rate and 
the energy requirements for bursters. \  
We also discuss two candidate power sources that could satisfy 
these requirements: magnetic energy and accretion energy. 
Several refinements of the basic HBM are 
investigated in \S \ref{sec-hbm-ext}, namely: gamma-ray intensities
that vary with angular position within the beam, GRB 
luminosity functions, and models with a significant subset of bursters
on bound orbits in the galactic halo.  
In \S \ref{sec-conclu} we give our conclusions 
and outline future observational tests.

Note that we consider possible physical mechanisms for halo GRBs, 
involving neutron stars with unusually strong magnetic fields, in 
\S 2.3, \S 2.4, \S 5, \S 6.3, and Appendix B. 
These sections could be read separately from the rest of the paper, 
since they might apply in a non-beamed model context (cf.~\S 7.2).

\section{The Halo Beaming Model: Physical Motivations}
\label{model}

\subsection{Model Assumptions}

We will assume that GRBs are emitted by HVNSs, 
which emanate from the galactic disk like a ``wind''
extending into the galactic corona (\cite{sm85}).
The HBM  does not require that the bursters or their peculiar 
velocities have any favored orientation in a galactic coordinate frame,
but it does invoke the physically-plausible condition 
that gamma-rays are produced only within 
a cone of angular radius $\phi_b$ about the star's magnetic axis 
$\pm \bf \overrightarrow \mu$. The magnetic axis is furthermore 
assumed to be roughly aligned (within $\sim 20^{\circ}$) 
with the stellar recoil velocity ${\bf V_r}$. \ 
We discuss the physics of such beaming and alignment in \S 
\ref{sec-phy-beaming}.

The particular version of HBM which we will analyze quantitatively in
\S \ref{sec-montecarlo} and \S \ref{sec-hbm-bat-pvo} below has the 
following simple properties:
[1] bursters are born at positions distributed like young Pop.~I stars
in the galactic disk; [2] with randomly directed recoils
$V_r = 1000$ km s$^{-1}$; [3] they emit GRBs at a constant rate, 
with [4] constant luminosity, and [5] the gamma ray emission 
is beamed parallel and anti-parallel to $\bf V_r$, 
within an angular radius $\phi_b$ that we will vary.

GRB beaming in a direction correlated with ${\bf V_r}$ 
makes the observable burst distribution comply with the BATSE 
dipole isotropy and the PVO brightness--distribution constraints, for 
the following reasons (see Figure \ref{fig-hbm-concept}).  
Since the HVNSs are freely--streaming out of the galaxy, 
their mean density diminishes with distance from the galactic center as 
$n\sim r^{-2}$.  \ However, the fraction of escaping bursters that are 
potentially {\it detectable} at Earth increases
in proportion to the transverse area of their beaming cones, $\sim r^2$. \ 
These two trends cancell, making the {\it effective} number of bursters
increase linearly with the sampled volume (i.e., an apparent
``constant density" of detectable stars) 
within a ``core radius'' $R_c \sim R_o/\phi_b$, where $R_o \sim 8.5$ 
kpc is a galactic disk dimension. In the HBM, this produces the 
the observed ``homogeneous'' distribution of bright GRBs found by the PVO 
experiment. \ Furthermore, since most bursters are undetectable when
they are at small $r$, the observable dipole anisotropy of 
GRB positions in the direction of the galactic center is greatly reduced.
At distances larger than $R_c$, {\em all} bursters are
detectable at Earth (or nearly all; see exceptions discussed in \S 3), 
and the $n\propto r^{-2}$ free-streaming fall-off of
burster density prevails, accounting for the ``boundedness'' ($\VVm < 0.5$) 
found by BATSE.

A more detailed illustration of the geometrical effect of burst beaming 
is given in Figure \ref{fig-zoa}. 
\ To understand this figure, consider for a moment
an idealized model in which all stars are born precisely at the 
galactic center (GC), and move out in random directions on straight--line
trajectories, each emitting GRBs into a cone of half-opening
angle $\phi_b$ around its velocity vector. Bursters lying outside the
circles in Figure \ref{fig-zoa}, or within the lens-shaped intersection
of circles between the GC and Earth,
are then the only ones that can be detected at Earth.
There is a large ``zone of avoidance'' (ZOA), within which all bursters are
invisible.  Figure \ref{fig-zoa}
actually shows only the 2-D cross-section of this ZOA. \ 
The true ZOA is the volume of revolution of the pictured shape about 
the line between Earth and the galactic center.  The boundary in Figure 
\ref{fig-zoa} at which stars just become observable (edge of the ZOA) 
is part of a circle with radius $D_{\rm sun} / \hbox{sin} \phi_b$, where 
$D_{\rm sun}=8.5$ kpc is the distance from Sun to the galactic center.

In the realistic situation, bursters are born throughout the galactic 
disk, with a peak birth rate at $\sim 4$--5 kpc from the center,
where the greatest concentration of Population I stars are located
(\cite{vdk87}).  Each birthplace in the disk then has its own ZOA, scaling 
up or down in linear size with the distance between
the birthplace and Earth.  One must add the weighted distributions of
detectable bursts together to get the total observable GRB distribution. 
There is no simple, analytic way to do this, so in what follows
we will use Monte Carlo methods.  We will also calculate realistic
trajectories in the galactic potential, rather than assuming 
straight lines.

\subsection{Physics of Beaming and Alignment} 
\label{sec-phy-beaming}

Gamma-ray bursts from high-$B$ environments tend to be beamed along field 
lines because of the transverse pair-creation opacity (e.g., 
\cite{rmb89}; \cite{hef90})
and because gamma rays produced by such mechanisms
as curvature radiation and Compton upscattering 
are strongly beamed along field lines (e.g., \cite{s86};
Dermer 1990 and references therein).  
Even locally--isotropic emissions are strongly beamed when they occur in 
a relativistic outflow channeled by a magnetic field (e.g., Yi 1993). 
Gamma emissions induced by sheared Alfv\'en waves in a neutron star 
magnetosphere can also be highly beamed (Melia \& Fatuzzo 1991; 
Fatuzzo \& Melia 1993).  Beaming obviates the $\gamma$--$\gamma$
opacity limit for GRBs (Krolik \& Pier 1991; Baring 1993; Fenimore
Epstein \& Ho 1993).

The HBM invokes the additional
(assumed) property that  burster recoil velocities are approximately 
aligned with magnetic dipole---and hence burst beaming---axi: 
$\overrightarrow V_r \parallel \overrightarrow \mu$.  \ This requires that 
the rotation axis $\overrightarrow \Omega$ is also roughly aligned 
with $\overrightarrow\mu$, at least to within the burst opening angle 
$\phi_b$. 

Is $\overrightarrow V_r \parallel \overrightarrow \mu$  
a plausible assumption?  Any recoil mechanism 
which imparts impulses to the stellar surface 
with a coherence time longer than the  rotation 
period of the star will tend to make 
$\overrightarrow V_r  \parallel \pm \overrightarrow \Omega$ 
because of ``rotational
averaging."  One such recoil mechanism is anisotropic neutrino emission
during the first $\sim 10$ s after formation (\S 2.4 below). \  If it is also 
true that $\overrightarrow \mu \parallel \pm \overrightarrow \Omega$ 
(at least to within the beaming angle $\phi_b$) then the 
alignment condition of the HBM is satisfied.   

Such near--alignment between  
$\overrightarrow \mu$  and $\pm \overrightarrow \Omega$ 
is expected if the neutron star magnetic field is generated by
large-scale dynamo action, as in familiar stellar and planetary dynamos. 
The ``magnetar" model, described below (\S 2.3), is one scenario
involving such a dynamo. 
Vacuum magnetic torques {\it increase} alignment (\cite{mg70}; \cite{dg70})
after the star spins down past the ``death line" 
at which magnetospheric currents are quenched, at 
age $\td \sim 10^5 \, (B_{dipole}/ 10^{14} \, \hbox{G)}^{-1}$ yrs 
(Chen \& Ruderman 1993). 
Before this point is reached, currents might exert counter-aligning
torques of comparable strength (e.g., McKinnon 1993 and references therein). 

Note that the GRB beaming angle $\phi_b$ is the sum of the intrinsic beaming 
angle due to gamma-ray opacity and radiative effects, 
and the r.m.s.~scatter in the angle between $\overrightarrow\Omega$
and $\overrightarrow\mu$.  \  Gamma production mechanisms such as 
curvature radiation and Compton upscattering could 
operate at significant altitudes in the magnetosphere, where field lines are
tilted with respect to ${\overrightarrow\mu}$, yielding larger 
values of $\phi_b$ than one would estimate under the assumption of near-polar 
gamma emission (as adopted by \cite{rmb89}; \cite{hef90}).
Values as large as $\phi_b \sim 20^{\scriptscriptstyle\circ}$ are 
possible.

\subsection{Magnetically-powered Bursts?}

The HBM might apply in a variety of different physical contexts. 
Here we will briefly discuss the ``magnetar" model for GRBs (DT92).
The magnetar idea has previously been invoked in models of 
soft gamma repeater bursts (\cite{td95}, hereafter TD95) 
as well as in models of  
continuous X-ray emissions from SGRs and from anomalous
X-ray pulsars like 1E2259$+$586 (\cite{td96}, TD96). 

Magnetars are hypothetical neutron stars which are born with 
dipole fields in excess
of $B_Q \equiv me^2 c^3/e \hbar = 4.4\times 10^{13}$ G. \ 
How might such stars form? \  Neutron stars are hot and convective 
during the first $\sim 10$ seconds after they form. 
They also undergo strong differential rotation (TD93).
Large-scale dynamo action might operate efficiently during this time 
for neutron stars with mean rotation periods below a threshold comparable
to the convective overturn time, 
as predicted by $\alpha$--$\Omega$ dynamo theory. 
There is evidence for such a threshold in magnetically-active 
main-sequence stars (e.g., Simon 1990).  {\it This threshold effect could 
give rise to  a bimodal population of neutron stars} 
(``pulsars and magnetars") with a factor $\gsim 10^2$ difference
in field strength (DT92). A more detailed explanation of
this magnetar formation hypothesis is given by Duncan \& Thompson (1996, 
DT96).   Seven distinct estimates of the field in the bursting
neutron star SGR 0526$-$66 seem to indicate $B > 10^{14}$ G 
(TD95).\footnote{Katz (1996) has raised some
interesting questions about theories of SGR burst emissions.  Our views
on these points are given in TD95. \ In particular, equation
(2) of TD95 explains why we believe that {\it plasma--confining} magnetic fields
$\gsim 10^{14}$ G are necessary in SGR 0526--66;
\ \S 3.2 of TD95 explains why we do not favor a model for SGR bursts
involving a pure pair plasma confined by a $\sim 10^{12}$ G field,
as favored by Katz; and \S 7.3.2 explains why we do not favor a 
hydrostatic model in which the magnetic Eddington limit operates without 
confining magnetic stresses, which Katz incorrectly attributes to us.
Another good argument against hydrostatic models, 
based on some novel physics, is given by Miller (1995).}

Magnetars spin down too rapidly to be observed as radio pulsars.
DT92 conjectured that GRBs are flare-like reconnection events in a galactic 
halo population of such strongly-magnetized neutron stars.  We will show 
in \S 4.5 that the available 
magnetic energy is sufficient to power observed GRBs,
in the context of the HBM (see also PRR). 

Magnetic reconnection is a theoretically 
advantageous energy source for GRBs, for several reasons.  
Gravitational and nuclear energy releases generally occur in bulk 
baryonic matter, which  has many degrees of freedom into which energy can 
thermalized and degraded via adiabatic expansion 
(e.g., \cite{ps93}). Flares in a neutron star 
magnetosphere, on the other hand, can be ``clean," exciting only photon 
and pair degrees of freedom to a first approximation.  Large scale 
electromotive forces induced by reconnection will generally accelerate 
pairs, and may produce hard, non-thermal spectra  via Compton upscattering, 
synchrotron emission, and curvature radiation (\cite{s86}). 
Indeed, GRBs have many qualitative similarities to stellar 
flares.  Both are transient energy releases, with chaotic
variability over a wide range of time scales; and both have spectrally-hard
non-thermal components (e.g., Murphy et al.~1993; Ramaty \& Mandzhavidze
1993). The similarities are so strong that 
a minor fraction of bursts in the BATSE catalog might actually be intense 
events from nearby, non-neutron, flare stars (Liang \& Li 1993). 
Stochastic avalanche models give a good fit to GRB time profiles
(Stern \& Svensson 1996) and to many properties of solar flares 
(Lu et al.~1993).

If the $V_r$--$B_{dipole}$ correlation observed in some samples of
radiopulsars (e.g., Cordes 1986; Stollman \& van den Heuvel 1986)
is applicable for all neutron stars, then the {\it mean} 
magnetar recoil velocity would be $\gsim 10$ times 
larger than the mean for radiopulsars
(but see Lorimer, Lyne and Anderson 1995). 
Several mechanisms predict unusually large recoils
for magnetars, sufficient to propel them into the galactic halo,
as explained in \S 2.4 below. 

Because of the dynamo origins of magnetar fields, one
expects that {\it initially} 
$\overrightarrow\Omega$, $\pm \overrightarrow V_r$ 
and $\pm \overrightarrow \mu$ are
roughly aligned in magnetars.\footnote{Magnetically-induced neutrino
starspots, as discussed in \S 2.4, 
will have coherence times $\tau_{spot} \gsim P_{rot}$ 
when the convective overturn time satisfies $\tau_{con} \gsim P_{rot}$,
thus ``rotational averaging" implies that
$\overrightarrow V_r  \parallel \pm \overrightarrow \Omega$ 
(where $\Omega = 2\pi/P_{rot}$); \ and 
$\overrightarrow \Omega \parallel \pm \overrightarrow \mu$  
because the dipole field is generated by an $\alpha$--$\Omega$ dynamo.}
As the star spins down, 
magnetospheric currents might drive some degree of misalignment; 
however, once the star spins down
past the ``death line" at time $\sim 10^5 \, 
(B_{dipole}/10^{14} \, \hbox{G})^{-1}$ yrs 
(\cite{cr93}), further spindown certainly enforces alignment 
(\cite{mg70}; \cite{dg70}), since the magnetostatic stellar 
distortion in magnetars is large enough to 
damp nutations of $\overrightarrow\Omega$ 
about $\overrightarrow\mu$ (\cite{g70}, DLT).
Incidentally, this is probably not true in old, spun-down ($P_{rot}>4$ s)
radiopulsars.
We conclude that $\overrightarrow V_r$  is likely to
be roughly aligned with $\pm \overrightarrow \mu$ in old magnetars, where 
$\pm \overrightarrow \mu$ is also the axis of beamed gamma 
emissions (DLT).\footnote{Fenimore, Klebesadel \& Laros (1996)
have suggested that an initial period of (spindown-induced)
misalignment occurs during the SGR phase, allowing bursts like
the 1979 March 5th event to be observed. Subsequent alignment by vacuum
magnetic torques occurs when classic GRB emissions begin.} 

This scenario for classic GRBs assumes that magnetars retain strong 
dipole magnetic fields.  Large-scale magnetic
instabilities that reduce $B_{dipole}$ (Flowers \& Ruderman 1979) might be 
suppressed by toroidal field components 
in the stellar interior, as expected for fields
generated via $\alpha$--$\Omega$ dynamos, 
or perhaps by other mechanisms (TD93 \S 14.2).  Note that 
a dipole field anchored in the stably--stratified liquid interior of 
a magnetar cannot be greatly distorted by 
spindown-induced crustal tectonic drift (Ruderman 1991)
because magnetic stresses dominate tensile stresses in the crust. 
This might differ markedly from 
the circumstance in ordinary radiopulsars (Ruderman 1991).
The interactions of vortex lines in the neutron superfluid with 
superconducting flux 
tubes are also probably irrelevant for {\it young} magnetars, because
the interior field is probably strong enough to suppress
superconductivity.  A young magnetar's field evolves predominantly via ambipolar
diffusion in the interior and Hall fracturing in the crust (TD96).
As the interior field decays, superconductivity will appear in the mantle
and perhaps the core; but this probably happens only after rapid spindown
has driven most superfluid vortex lines out of the interior. Further 
discussion of magnetar evolution is given by PRR.

\subsection{ Neutrino Magnetic Recoils}

A dipole anisotropy of only $\sim 0.03$ in the neutrino emission
from a young, hot neutron star would impart a recoil velocity 
$\sim 1000$ km s$^{-1}$ to the star (Chugai 1984).
Here we briefly review mechanisms whereby neutrino anisotropies could be 
{\it magnetically} induced.  
Purely hydrodynamic (non-magnetic) mechanisms for producing 
neutron star recoils have been proposed by Janka \& M\"uller (1994), 
Shimizu, Yamada \& Sato (1994), Burrows \& Hayes (1996), and in 
references quoted therein. 

 A strong magnetic field affects neutrino emissions from the 
beta processes $ n \rightarrow p^+ \ e^- \ \nu$ and
$p^+ \ e^- \rightarrow n \ \bar{\nu}$, as calculated by 
Dorofeev, Rodionov \& Ternov (1985), and from neutrino scattering 
processes as calculated by Vilenkin (1995).  Dorofeev et al. 
and Vilenkin furthermore estimated the neutron star recoils 
resulting from these processes, in the uniform-field 
idealization.  This is a macroscopic manifestion of parity-nonconservation
in the weak interactions.\footnote{ Another neutrino magnetic recoil 
estimate, by Bisnovatyi-Kogan (1996), makes several assumptions:
(1) the neutron star magnetic field undergoes simple linear winding 
driven only by differential rotation (i.e., ignoring all mixing motions); 
(2) the winded field is largely cancelled by a postulated 
pre-existing toroidal field in one hemisphere only; (3) the 
free neutron decay rate in a magnetic field applies.  These assumptions
are questionable.  In particular,  
realistic estimates of the beta neutrino emissivity within a nascent
neutron star should take into account the presence of magnetized, 
degenerate, relativistic electrons (Dorofeev, Rodionov \& Ternov 1985). 
Dorofeev et al.'s work was itself an improvement  
on Chugai's pioneering calculation, which did not treat the energy levels
of the electrons in a $B\gg B_Q$ field correctly.} 

In a realistically non-uniform
magnetic field, the back-reaction of magnetic stresses on convective energy 
flow, along with the neutrino opacity variations outlined above, would 
give rise to ``neutrino starspots," analogous to sunspots, 
and thus produce neutron star recoils of a magnitude estimated in DT92 and 
\S 13 of TD93. 

These calculations are based on standard Weinberg-Salam weak
interaction theory.  If neutrinos have mass, on the other hand, 
a strong magnetic field will shift resonant flavor-changing  
(MSW) oscillations, thereby also inducing an anisotropy in the emergent 
neutrino flux from a nascent neutron star (Kusenko \& Segr\`e 1996). 

All of these mechanisms produce recoils which vary 
directly---usually linearly---with the mean 
magnetic field strength: $V_r \propto B$.  \ Thus if any of these mechanisms 
dominate, one would expect much larger recoils for magnetars than for pulsars.  
Several {\it non-neutrino} recoil mechanisms which also might operate 
more efficiently in magnetars than in pulsars were discussed by 
DT92.\footnote{A star with an oblique, off-center dipole field experiences
a {\it spindown-driven} recoil (Harrison \& Tadmaru 1975), but this is unlikely 
to produce velocities as large as $\sim 1000$ km s$^{-1}$, as explained
in DT92.}
 
To estimate neutrino magnetic recoils realistically, one must know 
the strength, coherence length and coherence time of 
the magnetic field during the epoch of maximum neutrino luminosity.
This is not possible at present.  However,  
if the magnetic field approaches equipartition with the free energy of
differential rotation and/or the convective fluid mixing---which 
is strongly in the MHD limit---then 
$B \gsim 10^{16}$ G at the neutrinosphere 
(TD93).  Such strong fields could be present when most of the neutrino 
energy is radiated away, even if the surface {\it dipole} field, 
which is frozen-in tens of seconds later by the onset of stable stratification 
in the liquid interior (Goldreich \& Reisenegger 1992), 
is smaller by a factor of $\sim 10$ or $\sim 30$. \  Recoils $V_r \gsim 10^3$
km s$^{-1}$ are plausible.

\section{Galactic Halo Model Results}
\label{sec-montecarlo}

By integrating a large number ($\sim 10^6$) of neutron star 
trajectories in a realistic galactic potential,
we have derived the sky distribution of HVNSs.
In the model described here (\S 3 and \S 4), all stars have 
$|{\bf V_r}| = 1000$ km s$^{-1}$; thus they move 
in nearly straight lines and eventually escape from the galaxy (but
see \S 6.3).

We used a Monte Carlo code developed by Li \& Dermer (1992), which 
in turn follows many of the prescriptions of Paczy\'nski (1990) and 
Hartmann, Epstein \& Woosley (1990).  
In particular, we adopt van der Kruit's (1987) model for the spatial 
distribution 
of young Pop.~I stars which spawn neutron stars in the galactic disk.  
The trajectories of $\sim 10^6$ neutron stars were numerically
integrated in a realistic galactic potential (Miyamoto \& Nagai 1975)
\footnote{Lamb, Bulik, \& Coppi (1996) have argued that the galactic 
potential  of Miyamoto \& Nagai (1975) implies an 
unrealistically extended disk which tends to `focus' neutron star orbits 
into the disk plane. We find, however, that this focusing
effect is negligible for {\it randomly} oriented initial recoil velocities
of magnitude $V_r\gsim 1000$ km s$^{-1}$.  \ This is because the disk is only a
very minor part of the total mass of the Galaxy.}
including a dark halo cutoff at radius $R_H = 70$ kpc.  
The escape velocity from the
center of this model potential is $V_{esc}\sim 600$ km s$^{-1}$.  
This increases only logarithmically with $R_H$, reaching $800$ km s$^{-1}$ 
for $R_H = 200$ kpc, thus our results are probably insensitive to $R_H$ 
over its plausible range (LD92).   Deviations from 
sphericity in the dark halo were neglected (but see PRR). 
We also do not include the potential of M31 since 
stars with $V_r \sim 10^3$ km s$^{-1}$ are negligibly
perturbed by M31 within the BATSE sampling depth we found of $<200$ kpc
(\S 4.1).

In Figure \ref{fig-ang-hbm} we show HBM numerical results for several angular
statistics, plotted as functions of sampling depth $D$ about the Earth.
That is, the figure shows cumulative angular statistics
for all detectable bursters at distances from Earth that are 
less than or equal to the value of $D$ on the horizontal axis. 
The topmost plot of Figure \ref{fig-ang-hbm} 
shows the galactocentric dipole moment 
$\cost$, where $\Theta$ is the angle between a burst and the galactic
center. The second plot shows the disk-like quadrupole, $\sinb - 1/3$, 
where $b$ is galactic latitude; 
$\sinb < 1/3$ implies that sources are concentrated toward the disk.
The third plot shows the galactocentric quadrupole, $\coss - 1/3$. \
The bottom plot of Figure \ref{fig-ang-hbm} shows $\VVm$, 
a statistic which is related to the slope of the cumulative 
log ${\cal N}$---log $P$ brightness distribution,
as explained below. 

Within each subplot of Figure \ref{fig-ang-hbm}, the 
various lines correspond to different beaming
angles $\phi_b$ as described in the figure caption.
We have cut off the $\phi_b= 10^\circ$ and $5^\circ$ curves
for $D \leq 10$ kpc because our Monte Carlo sampling statistics at 
smaller $D$ are too poor.  For example, the number of 
detectable stars at $D\le 3$ kpc in our Monte Carlo model is only 
$\sim 55$ for $\phi_b = 5^{\circ}$,  whereas it is $\sim 10^4$  for in the
unbeamed case ($\phi_b = 90^{\circ}$).

Increasing the sampling depth $D$ (moving to the right in Figure 
\ref{fig-ang-hbm}) is tantamount to including fainter and fainter bursts.  
Eventually the BATSE sampling depth 
is reached; at this point, if the model is to fit
observations, all the plotted statistics must simultaneously match 
BATSE values to within observational uncertainty.  
The most recent published BATSE results\footnote{ Results from the 
third BATSE catalog for 1122 bursts (Meegan et al.~1996) are as follows:  
$\cost_{data} = - \, 0.002 \pm 0.017$. 
There is a concurrent anisotropy in the BATSE sky coverage of  
$\cost_{sky} = - 0.013$.  \  
Thus the best measure of the burst population dipole moment
is $\cost = \cost_{data} - \cost_{sky} = + 0.011 \pm 0.017$. \
Similarly, $\langle$sin$^2 b \rangle - 1/3 = 0.002 \pm 0.009$, and \
for 657 bursts, $\VVm = 0.33 \pm 0.01$. }
are plotted in Figure \ref{fig-ang-hbm} at a value $D = 180$ kpc.  
Before discussing these results, we must explain the significance of $\VVm$.

The $V/V_{max}$ statistic 
was invented for the study of the quasars (Schmidt 1968),
and was first applied to GRBs by Schmidt, Higdon \& Hueter (1988).
For scintillation counter experiments, $\VVm$ is the average of
$(C/C_{min})^{-3/2}$ over all bursts, where $C$ is the peak 
counts (in a set time interval) and $C_{min}$ is the threshold for
detection.  $C_{min}$ can vary with background noise and other effects
(e.g., ``overwrites"); however, for a source population that is uniformly 
distributed in static Euclidean space, 
$\VVm$ is equal to 0.5 regardless of how the threshold varies; 
and $\VVm < 0.5$ indicates that the density of bursters diminishes 
with distance, for standard-candle sources. 
Because $C_{min}$ is variable, the $\VVm$ statistic is useful 
in the study of inhomogeneous data sets {\it only} when trying to answer the 
yes/no question: ``Is the observed brightness distribution consistent 
with a uniform density of sources distributed in Euclidean space?"  
(e.g., Band 1992; Petrosian 1993).

In Figure \ref{fig-ang-hbm}, we use $\VVm$ in a
purely illustrative way.  Our model values of $\VVm$ are ideal values 
that would be found by an instrument with a uniform detection 
threshold, i.e., no variations in the noise or the threshold settings.
Since the $C_{min}$ values in the BATSE catalog are not highly variable
(Meegan et al.~1996) preliminary comparisons with BATSE, as in the 
bottom panel of Figure \ref{fig-ang-hbm}, will not lead us astray.
However, the most accurate way to compare burst observations with 
theory is to fit the 
observed {\it distribution} of peak photon fluxes to Monte Carlo models which 
have been realistically filtered for detection incompleteness (e.g., Lubin \&
Wijers 1993).
This is what we do in our actual statistical comparisons 
of the HBM with the BATSE catalog (\S 4). 

Note that the model curves for angular statistics in 
Figure \ref{fig-ang-hbm} 
implicitly assume a detector with uniform sky coverage;
i.e., we have idealized that the detector is equally capable of
detecting bursts from any location on the celestial sphere.
We have corrected for the imperfect sky coverage of BATSE by shifting
the {\it data points} in Figure \ref{fig-ang-hbm} 
appropriately (see footnote 9). 
In \S \ref{sec-hbm-bat-pvo} 
we will take the opposite approach: using the raw BATSE data  
and filtering the Monte Carlo model to take into account BATSE's 
imperfect sky coverage.

What can be learned from Figure \ref{fig-ang-hbm}? \  
Beaming evidently increases $\VVm$ to nearly 0.5 for nearby
bursters, $D\sim 30$ kpc. This can make the model satisfy PVO constraints,
as we show quantitatively below.  Beaming also evidently reduces $\cost$. \ 
The $\phi_b = 20^\circ$ case fits BATSE observations 
over an appreciable range of sampling depths $D> 100$ kpc.  

Note that the largest deviations from isotropy in Figure \ref{fig-ang-hbm}
occur for bright (i.e., small-$D$) subsets of the
observable bursts.  A distinctive signature of the HBM 
is that $\coss - 1/3$ is positive for bright bursts. 
This can be understood as follows. 
The (bright) bursters which become visible 
at Earth before reaching the remote galactic halo are the ones
which happen to be born with beaming axis $\overrightarrow \mu$ 
(and hence, with $\overrightarrow V_r$) pointed approximately toward 
or away from Earth (to within $\sim \phi_b$).  Because most bursters
are born within the Solar circle in the galactic disk, the
brightest ones tend to be seen in that direction and toward
the galactic anticenter, if they have already moved past the Earth 
on their way out of the galaxy.  Thus $\coss > 1/3$ for
bright bursts, while the dipole moment $\cost$ remains small.

All burst sources can be seen at distances $D\gg R_o/\phi_b$ 
in the $\phi_b = \{ 90^\circ, 40^\circ$ and $20^\circ \}$ cases of 
Figure \ref{fig-ang-hbm}, 
thus in a first approximation, one would expect $\sinb$ to 
reach the isotropic value of 1/3 asymptotically at large $D$.  
In fact, a mild degree of disklike quadrupole anisotropy, 
$\sinb = 0.31$, remains in these models at large $D$. \ 
This is due to the  
galactic disk rotation (with velocity $V_\Omega \approx 220$
km s$^{-1}$), which makes bursters tend to stream out at lower galactic 
latitudes, diminishing the density of bursters near the galactic
poles compared to the equator at a given distance $r$ from the galactic 
center (LD92). This effect is not dependent on beaming.
It depends only on the ratio $V_\Omega/V_r$, with $\sinb \to 1/3$
at large $D$ as $V_\Omega /V_r \to 0$.    

Galactic rotation also causes the total initial velocity 
vector $\vrec + \vgal$ to be inclined by as much as as 
\begin{equation}
\label{eq-tilt-ang}
\phivmax \equiv \hbox{arctan}(V_\Omega/V_r) \approx 12^\circ
\ \left({V_r \over 1000 \, \hbox{km s}^{-1}}\right)^{-1}
\end{equation}
with respect to the axis of GRB emission, $\vrec/V_r$.  \ 
Thus if $\phi_b < \phivmax$, some stars never become visible from Earth.
Since this happens preferentially at high galactic latitudes, 
$\sinb$ is always substantially less than 1/3 in models with 
$\phi_b < \phivmax$. \ 
For example, in the case $\phi_b = 10^\circ$, $\sinb$ never exceeds
0.25.  Only when $\phi_b > \phivmax$ 
are all stars visible at large $D$, causing $\sinb$ to asymptote to the 
unbeamed, large--$D$ value of 0.31. 

When $\phi_b < \phivmax$ most stars  
eventually become undetectable at Earth, but only after passing through 
a phase when they can be seen,
namely, when $D < 8.5 \, \hbox{kpc} / [\hbox{tan} \, (\phivmax - \phi_b)] 
\approx 150$ kpc.
This is evident in the quadrupole moment plots of 
Figure \ref{fig-ang-hbm}.

\def\degrees{{\scriptscriptstyle \circ}}
This leads us to consider bounds on the range of acceptable 
beaming angles $\phi_b$. \ The effective ``core radius" in the HBM, 
is  $R_c \simeq R_o/\phi_b \simeq 24 \, (\phi_b/20^\degrees )^{-1}$ kpc.
This value is confirmed by our numerical results if 
we define $R_c$ as the sampling depth at which $\VVm = 0.46$  
close to the value found by PVO (Fenimore et al.~1993).
Thus by equation (\ref{eq-tilt-ang}),
\begin{equation}
\label{eq-cradius}
R_c \simeq 40 \, V_3 \ \left({\phi_b\over \phivmax}\right)^{-1} \
\hbox{kpc}, 
\end{equation}
where $V_3 \equiv V_r / 10^3$ km s$^{-1}$.  \ 
Since BATSE bounds on $\sinb$ imply $(\phi_b /\phivmax)>1$, the HBM cannot
have a core radius larger than $\sim 40 \, V_3$ kpc.  The {\it minimum} 
core radius that is likely to fit the data is $R_c\sim 16$ kpc 
(Paczy\'nski 1991), thus the range of allowed $\phi_b$ is roughly
\begin{equation}
\label{eq-phib}
12^\circ \, V_3^{-1} < \phi_b \le 30^\circ \ . 
\end{equation}
This agrees with Figure \ref{fig-ang-hbm}, 
where $\phi_b = 20^\circ$ give the best fit
to BATSE data.

Although Fig.~3 is for $V_r = 1000$ km s$^{-1}$, 
note that better fits to the BATSE data can be obtained 
when $V_r > 1000$ km s$^{-1}$. This reduces the effect of galactic
rotation on the burster distribution, causing $\sinb$ to converge
toward 1/3 at the faint end, and it  
also widens the acceptable range of $\phi_b$ (eq.~[\ref{eq-phib}]).   
Some soft gamma repeaters have recoils as large as 
$V_r \sim 2000$ km s$^{-1}$ (Appendix A).  

\section{Model Comparisons with the 3B Catalog and PVO}
\label{sec-hbm-bat-pvo}

We now make detailed comparisons of the HBM with the 
3rd BATSE catalog (\cite{mee96}), which we will refer to simply as ``3B". 
Because the largest deviations of angular moments  
from isotropy in the HBM occur within
bright subsets of bursts (small $D$ in Figure \ref{fig-ang-hbm}), 
we will explicitly consider the cumulative angular statistics 
as a function of observed peak photon fluxes, $P$.   
This is the basic approach that we previously applied to the 
first and second BATSE catalogs (LDT; LD96a).  A preliminary 
discussion of 3B comparisons is given by LD96b; see also Bulik
\& Lamb (1996) for a complimentary statistical analysis. 

\subsection{The Log ${\cal N}$--Log $P$ Distribution}

There are several issues that must be faced when doing statistical
analyses of the 3B catalog.  There are data gaps in  
in many bursts, and for many bursts the
associated trigger efficiency (the probability that BATSE will detect 
a burst, as a function of peak flux) is less than one.
\ Under these circumstances, we use
the following criteria for selecting bursts from BATSE and PVO:
{\bf (1)} Overwriters in 3B are not included.
{\bf (2)} Only bursts from 3B with $P_{256} \geq 1$ photons cm$^{-2}$ s$^{-1}$, 
whose trigger efficiency presumably is $\geq 1$, are used.
Similarly, bursts from PVO with $P_{250} \geq 25$ photons cm$^{-2}$ s$^{-1}$
are used.
{\bf (3)} Only bursts for which $({C_{\rm max}/ C_{\rm min}})_{256} \ge 1$
or $({C_{\rm max}/C_{\rm min}})_{1024}\ge 1$, or both, are used.
This applies to both 3B and PVO.

This counting criterion reduces the number of BATSE bursts used in 
our study of the log ${\cal N}$--log $P$ distribution  
to 241 out of the total number 1122 in 3B.\footnote{
We will use many more 3B bursts in our study of the GRB angular 
distribution, as described below.  Note that
a significant fraction of 3B bursts do not have $C_{\rm max}/ C_{\rm min}$
available, so we had to drop these bursts from the start.}
This, however, also allows us to construct a relatively homogeneous
{\it merged} catalog with the PVO data, since the PVO bursts are 
selected by very similar criteria (Fenimore et al.~1993).
Thus we add 181 PVO bursts to our analysis of the 
log ${\cal N}$---log $P$ distribution, making a total of 422.  
More importantly, by adding PVO
we have increased the range of peak fluxes which we 
can fit by a factor of about 20.

We have also done analyses by selecting bursts according to  
criteria {\bf (1)} and {\bf (2)}, and ignoring criterion {\bf (3)}. This is
somewhat justified since the BATSE bursts
for which $P_{256}$ and $C_{\rm max}/ C_{\rm min}$ are available all
have $({C_{\rm max}/ C_{\rm min}})_{256} \ge 1$
or $({C_{\rm max}/C_{\rm min}})_{1024}\ge 1$ when
$P_{256} \geq 1$ photons cm$^{-2}$ s$^{-1}$. This gives us a 
larger sample, with 437 bursts from BATSE and 198 from PVO, thus
making a total of 635. The fitting parameters we obtained from this
sample, especially the peak luminosity, are very similar
to the values we obtained from fitting the 422 sample, however. Thus we will
only present the results from the sample of 422.

The combined PVO and BATSE intensity distribution is
shown in Figure \ref{fig-lognp-pbh}. 
The HBM result for $\phi_b = 20^\circ$ and $V_r = 1000$ km s$^{-1}$ is 
also shown.  We have used two free parameters in fitting these data 
to the HBM: the absolute number normalization
(i.e., the number of GRBs per year) and the intrinsic peak photon emission
rate $\dot N$.  \ We find a best-fit ``standard candle" value of 
\begin{equation}
\label{eq-pkph}
\dot {N} \simeq 9.45 \times 10^{46} 
\ \left({1 - \cos \phi_b \over 0.06}\right) \ 
\hbox{photons s}^{-1}
\end{equation}
This fit has a minimum $\chi^2 \simeq 6.9$ with 
12 degrees of freedom.
The 1--$\sigma$ confidence region of $\dot N$, estimated as the range over
which  $\chi^2$ increases by 2.3, is $8.81 - 11.05 \times 10^{46}$ photons
s$^{-1}$. The corresponding peak luminosity is 
\begin{equation}
\label{eq-pkeng}
\dot E \approx 1.5\times 10^{40} 
\ \left({1 - \cos \phi_b \over 0.06}\right) \ 
\hbox{ergs s}^{-1} 
\end{equation}
if the average photon energy is $\sim 100$ keV. 
The corresponding BATSE sampling depth is $D_{\rm bat} \approx 182$ kpc 
for a limiting flux of $\sim 0.3$ photons cm$^{-2}$ s$^{-1}$. 

We have also fit the model to 3B alone, ignoring PVO constraints. 
This is shown in Figure \ref{fig-lognp-bh}.
The best-fit peak luminosity turns out to be 
\begin{equation}
\label{eq-pkph-bat}
\dot N = 6.9 \times 10^{46}
\ \left({1 - \cos \phi_b \over 0.06}\right) \ 
\hbox{photons s}^{-1}, \ \ \ \ 
\ \hbox{BATSE alone} 
\end{equation}
with a 1-$\sigma$
confidence region ranging from $4.7 - 10.4 \times 10^{46}$ photons s$^{-1}$. 
This is smaller than the number we got using the combined PVO 
and BATSE data, although the confidence regions do overlap.

\subsection{The Angular Distributions}

In order to compare the HBM with the observed burst angular
distributions, we have filtered our model results
through BATSE's sky exposure table. 
All bursts from 3B with tabulated values of $P_{256}$ are used in 
this comparison except overwriters,
giving us 814 bursts in total.

In Figure \ref{fig-ang-hbm-3b}, we plot measures of angular 
$($an$)$isotropy in galactic-based coordinates for bursts 
from the 3B catalog in which the peak photon flux 
($P_{256}$) is greater than or equal to
a given value, shown on the horizontal axis. Here 
$\phi_b= 20^{\circ}$, and we have 
used the best-fit photon luminosity $\dot N$ of equation \ref{eq-pkph}. 

The HBM curves (solid lines) make the distinctive 
prediction that $\langle \cos^2 \Theta \rangle > 1/3$, 
as explained in \S \ref{sec-montecarlo}. \
Dashed lines show model predictions if gamma-ray bursts 
are isotropic on the sky at all intensity levels, as in the
cosmological burster hypothesis. 
Note that both model curves (HBM and isotropy) 
are corrected for the incomplete BATSE sky coverage.
The numerical (Monte Carlo) uncertainties on the HBM curves
are very small on the scale of this graph:
comparable to the width of the thick solid lines.

Filled dots with error bars in Figure \ref{fig-ang-hbm-3b}
are BATSE cumulative data points
with 1--$\sigma$ statistical uncertainties evaluated according to
\begin{equation}
\label{eq-errors}
\sigma (\langle X \rangle) = N^{-1/2} [ \langle X^2 \rangle - 
\langle X \rangle^2 ]^{1/2} 
\end{equation}
(e.g., Briggs 1993)
where $X \equiv \cos\Theta,~\sin^2 b~{\rm or}~\cos^2 \Theta$, and
the averages $\langle X \rangle$ and $\langle X^2 \rangle$ 
are taken in the HBM. 
These error bars, appropriate for comparing observations 
with the HBM (solid lines), are slightly larger than the error 
bars that would be appropriate for comparing with the 
isotropic model (dashed lines).\footnote{ Briggs et al.~(1996) 
have studied the angular distributions 
of 1005 GRBs, putting constraints on galactic models, including the HBM.
They found a deviation  of $3.1 \, \sigma$
between the HBM and BATSE values of $\coss$ for the brightest 1/8
of their sample, which is larger than we find here. 
\ Our analysis differs from Briggs et al.~in two ways. 
First, a new intrinsic peak luminosity was found from fitting 
3B/PVO (\S 4.1), which is slightly larger than the luminosity 
that DLT found from fitting 1B/PVO, as used by Briggs et al. 
Second, our statistical uncertainties 
come from equation (\ref{eq-errors}) where the averages
$\coss$ and $\cosq$ are evaluted in the 
HBM rather than in isotropic models.  Isotropic models underestimate
the relevant uncertainties by more than $10\%$. }

Figure \ref{fig-ang-hbm-3b} shows that the 3B catalog is statistically 
consistent with both isotropy and the HBM 
at the 2--$\sigma$ level or better, 
for all angular statistics and at all peak photon fluxes. \ 
Nevertheless, isotropy fits the observations better than the HBM in
every graph, unlike in our earlier studies based on 1B and 2B 
(LDT; LD96a). At the bright end, the HBM offers several clear 
predictions which may be tested as BATSE accumulates more data.

\subsection{Short \& Long Burst Subclasses in the HBM}

When the numbers of BATSE bursts are plotted as a function of 
burst duration (i.e., a histogram), they show a statistically significant
two-peaked distribution (Kouvelioutou et al.~1993; Koshut et al.~1996; Meegan
et al.~1996).  This suggests that GRBs can be physically grouped into
two classes. In this paper we have analyzed the distribution of short and 
long bursts together, using the photon peak flux on a 0.256 s
time scale, $P_{256}$, as a brightness measure in order to make  
a homogeneous merged set of data from BATSE and PVO. \ 
However, in two earlier papers (LDT and LD96a), we studied the BATSE 
duration subclasses independentally, and compared them to HBM predictions. 
This was useful for exploring the possibility that the HBM, or indeed
any galactic halo model, might apply to one subclass only. 
In these studies we used $P_{1024}$ 
for the long bursts ($T_{90} > 2$ s) and
$P_{64}$ for the short bursts ($T_{90} < 2$ s).
The peak luminosity obtained for the 2B long bursts 
alone was $(4.0\pm1.1) \times 10^{46}$ photons s$^{-1}$ (LDT), consistent
with the values given in eqn.~(\ref{eq-pkph-bat}). \
Our plots of GRB angular statistics as a function of burst brightness
for the two subclasses (analogous to Figure 6) 
showed that the BATSE data was fit by
the HBM and isotropic (cosmological) models almost equally well.
No statistically significant deviations from isotropy or from the HBM were
found in either subclass at any peak flux level.

\section{Burster Energy Budgets and Repetition Rates}
\label{sec-eng-rep}

If halo GRB sources are observed as soft gamma repeaters (SGRs) 
when they are young, then a rough lower bound on the galactic 
formation rate of bursters is $\Gamma > 10^{-4}$ yr$^{-1}$.  \ 
This is based on the fact that 3  SGRs have been detected in likely association
with supernova remnants of age $\sim 1 \times 10^4$ years.  
Note that SGR activity is episodic (e.g., SGR 0526--66 was observed
to burst only during the years 1979--1983; Norris et al.~1991),
and only a small fraction of Type II supernova remnants have confirmed
radiopulsar (non-SGR\footnote{The $P_{rot} = 8.0$ s rotation 
period and other properties of SGR 0526--66 clearly indicate that 
SGRs are, as a class, distinct from rotation-powered pulsars. 
Only one SGR, 1806--20, is embedded in a radio plerion, 
and this has a steeper spectrum than a pulsar plerion (Kulkarni et al.~1994). 
This plerion might be magnetically--powered, as explained in Appendix A.})
counterparts (Frail 1997; Frail \& Moffett 1993), thus there might exist many 
heretofore undetected SGRs of ages $\lsim 10^4$ yrs in the galaxy.
Parameterizing $\Gamma_3 \equiv (\Gamma / 10^{-3}$ yr$^{-1}$), we find:
\begin{equation}
\label{eq-brate}
0.1 < \Gamma_3 < 10 , 
\end{equation}
where the upper bound comes from the fact that the rate 
of burster (SGR) formation cannot exceed the total galactic supernova 
rate.
 
If the BATSE sampling depth is $D= 100 \ D_{100}$ kpc and if all bursters 
acquire recoils of magnitude $\sim 1000 \ V_3$ km s$^{-1}$ (cf.~Appendix A), 
then the time for a burster to drift out to the BATSE sampling depth is
$t_D = 10^8 \, D_{100} \, V_3^{-1}$ yrs.  
The total number of bursters in the volume sampled by BATSE 
is then
\begin{equation}
{\cal N}_D = \Gamma \ t_D = 10^5 \, D_{100} \, V_3^{-1} \, \Gamma_3. 
\label{eq-nhalo}
\end{equation}

It is worthwhile emphasizing the meaning of this equation.
Given the observed SGR formation rate [eq.~(\ref{eq-brate})] 
and their observed recoils (Appendix A) we can conclude
on a purely empirical basis that 
{\it there exists a galactic halo of $> 10^4$ aged SGRs 
within the distance} 100 kpc (DT92).
The number of old SGRs in the halo might be orders of magnitude higher if 
$\Gamma_3 > 0.1$, or if a significant fraction of SGRs 
remain bound to the galaxy (eq.~[\ref{eq-ratiobound}] below). 
SGRs are known to emit multiple soft gamma-ray bursts of energy
$\sim 10^{41}$ ergs when they are young; if they continue to emit bursts 
with similar energies but harder spectra as they age, then 
they constitute halo GRB sources (DT92). 

The (extrapolated) full-sky rate of bursts that are bright enough to
be detected by BATSE is $\sim 800$ yr$^{-1}$ 
(Meegan et al.~1992).  If emitted by aged
SGRs,  then the mean repetition time for bursts from a given star 
is\footnote{ In the HBM, the beaming of bursts reduces 
the number of {\it observable} bursters by only a mild factor 
$f \approx 1- (2/3) R_o / (\phi_b D) \approx 1-0.16 \ D_{100}^{-1} 
(\phi_b/ 20^\circ)^{-1} \approx 0.84$, where $R_o \sim 8.5$ kpc is a
galactic disk dimension. This (unimportant) factor is included 
eq.~(10).}  
\begin{equation}
\label{eq-repe}
\taugrb \simeq 110 \ \Gamma_3 \ V_3^{-1} \ D_{100} \ \hbox{yrs}
\end{equation}
This implies that evidence for repetition might be difficult to find
in presently available data.

The intrinsic luminosity of a GRB near the BATSE detection threshold of 
$F_{\rm \scriptscriptstyle BATSE} \simeq 1\times 10^{-7}$ 
erg cm$^{-2}$ s$^{-1}$ (Meegan et al.~1992) is 
\begin{equation}
\label{eq-lum}
\dot E_{\rm GRB} \simeq 1\times 10^{41} \ D_{100}^2 \  
\left( { \Delta \Omega \over  4 \pi}\right) \ \hbox{erg},
\end{equation}
where $\Delta \Omega$ is the solid angle of the burst 
emissions.  This agrees with the luminosity quoted 
in equation~(\ref{eq-pkeng}), for a best--fit BATSE 
sampling depth $D_{100} = 1.82$ and $\phi_b = 20^\circ$. \ 
The effective duration of BATSE GRBs is $\sim 10$ s, as found from
the mean ratio of fluence to peak flux (Lee \& Petrosian 1996).
Thus the burst energy is 
\begin{equation}
\label{eq-eng}
E_{\rm GRB} \simeq 1\times 10^{42} \ D_{100}^2 \   
\left( { \Delta \Omega \over  4 \pi}\right) \ \hbox{erg}.
\end{equation}

In its observable (by BATSE) lifetime of 
$t_D = 10^8 \, D_{100} \, V_3^{-1}$ yrs, each star emits  
a number of bursts ${\cal N}_B = t_D/\taugrb =10^6 \ \Gamma_3^{-1}$.  The total
energy required from each burster is then 
$E_{\rm tot} = {\cal N}_B \, E_{\rm GRB}$, or
\begin{equation}
\label{eq-teng}
E_{\rm tot} \simeq 3\times 10^{46} \ D_{100}^2 \ \Gamma_3^{-1} \
\left({1 - \hbox{cos} \phi_b \over 0.06 } \right) \
\hbox{erg}. 
\end{equation}
This is the basic energy requirement for burst sources in the HBM. \
Note that in the small-angle limit\footnote{In the remainder of \S 5 and
in \S 6.3 we will quote many results with a scaling factor 
$(\phi_b/20^\circ)^2$.  These equations are given implicitly in the 
$\phi_b \ll 1$ limit. However, all these results can be generalized to
any $\phi_b$ by using eq.~(14).  In particular, a magnetar model with
isotropic ($\phi_b = \pi/2$) bursters could agree with BATSE and PVO 
data if the initial onset of bursting were delayed (LD92; Bulik \& Lamb
1996; Appendix B,
eq.~[B7]).  Energy budget constraints for this scenario are given by 
dropping the $(\phi_b/20^\circ)^2$ factor in eq.~(17) 
and multiplying by a numerical factor 
$(0.06)^{-1}= 16.6$.  These same isotropic results would apply if bursts 
were beamed, but in random directions; for example, if the alignment scenarios 
of \S 2.2 did not operate.   Although the energy per burst would then be 
reduced by $(\Delta \Omega/4\pi)$, one would need $(\Delta \Omega/4\pi)^{-1}$
more bursts to produce the same rate of observable events; these factors
cancell in the energy budget (Hartmann \& Narayan 1996).}
\begin{equation}
\label{small-angle}
\left({1 - \hbox{cos} \phi_b \over 0.06 } \right) \approx 
\left({\phi_b \over 20^\circ}\right)^2, \qquad\qquad  
\phi_b\ll 1.
\end{equation}

\smallskip
Rotational energy,
the power source for ordinary pulsars, is probably not capable 
of providing this much energy for GRBs
(LD92; Hartmann \& Narayan 1996). Furthermore, there is evidence
that SGRs are slow rotators ($P_{rot} = 8.0$ s for SGR 0526--66). 
We now discuss more promising energy sources.
 
\subsection{Magnetic Energy}
 
A neutron star with interior flux density 
$B = 10^{16} \, B_{16}$ G, idealized as uniform throughout the star, has
magnetic energy 
\begin{equation}
\label{eq-beng}
E_B = B^2 R^3/6 = 2 \times 10^{49} \, B_{16}^2 \ \hbox{erg}.
\end{equation}
\noindent This is a plausible estimate for the magnetic energy of a 
young neutron star with an internal toroidal field generated by a
saturating $\alpha$--$\Omega$ dynamo (DT92; TD93).  
It exceeds by a factor $\sim 10^3 \, \Gamma_3$ the GRB energy requirement
of eq.~(\ref{eq-teng}).  
The transition to superconductivity occurs at
$B< B_c \sim 3 \times 10^{15}$ G.  \ 
When $B < B_c$, magnetic flux is concentrated
by supercurrents into discrete flux tubes, and the energy is (PRR):
\begin{equation}
\label{eq-bbeng}
E_B = B B_c R^3/6 = 5 \times 10^{47} \, B_{15} \ \hbox{erg}.
\end{equation}
As a magnetar field evolves and decays it will enter the regime of eq.~(16),
even if it is in the eq.~(15) regime initially.

Equations (\ref{eq-teng}) and (\ref{eq-bbeng}) together imply
that magnetic energy can suffice to produce all GRBs in the HBM
if the burster internal magnetic field satisfies
\begin{equation}
\label{eq-breq}
B > 6 \times 10^{13} \ D_{100}^2 \ \Gamma_3^{-1} \ 
\left({\phi_b \over 20^\circ} \right)^2 \ \hbox{G}. 
\end{equation}
This criterion is met, and in fact exceeded by a 
factor $\sim (10$--$10^3) \, \Gamma_3$, in magnetar formation scenarios 
(DT96) and in the magnetar model of SGRs (TD95, TD96).  

Regardless of the energy source for bursts, if 
the energy release occurs {\it inside} a neutron star and is communicated
outward by material and Alfv\`en waves (e.g., Blaes et al.~1989), 
then very strong magnetic fields, $B>B_Q$ are probably necessary.
Weaker fields would allow vibrational crumbling of a halo neutron star's 
crust and thereby produce an X-ray excess (Appendix B).

\subsection{Accretion Energy}

The sudden accretion of a planetesimal (asteroid or comet) with mass
\begin{equation}
\label{eq-mplanet}
M_{p} = 5\times 10^{21} \ D_{100}^2 \ \epsilon_\gamma^{-1} 
\left( { \Delta \Omega \over  4 \pi}\right) \ \hbox{gm}
\end{equation}
onto a neutron star would release the energy of a halo GRB, 
where $\epsilon_\gamma$ is the efficiency of
gamma ray production.   
This is a scaled-up version of the GRB theory
of Harwit \& Salpeter (1973) and Colgate \& Petchek (1981). 

To power all $10^6 \ \Gamma_3^{-1}$ bursts in a burster's observable
lifetime, the total mass of accreted planetesimals is 
\begin{equation}
\label{eq-tplanet}
M_{\rm tot} = 5\times 10^{27} \ \Gamma_3^{-1} \ D_{100}^2 \ 
\epsilon_\gamma^{-1} \left( { \Delta \Omega \over  4 \pi}\right) \ 
\hbox{gm},
\end{equation}
similar to the mass of the Earth.  These planetesimals presumably
condense from a gaseous disk.  Such a disk 
($M_{\rm disk} > M_{\rm tot}$) could be captured by a HVNS
when it overtakes clumpy supernova ejecta (Woosley \&
Herant 1996) or experiences a recoil directed toward a stellar
companion (Colgate \& Leonard 1994).  Alternatively, a disk could form
when a rapidly-rotating proto-neutron star sheds excess angular momentum
(e.g., Thompson \& Duncan 1994).  In this case, the intial radius of
the retained disk could not exceed 
$R_{i, \rm disk} \sim (GM_\star/V_r^2) \sim 2\times 10^5 \ V_3^{-2}$ km;
but it would subsequently experience viscous spreading.

It is not obvious how the beaming and alignment requirements of the
HBM are met in such an accretion-powered scenario, although the
physics is complex enough that a mechanism might be identified.
Alternatively, GRBs could be unbeamed, 
or beamed in random directions, but with a delayed onset of bursting (LD92).
Colgate \& Leonard (1994) estimated that a time interval $\sim 10^7$
years, as required to fit BATSE data, occurs before planetesimals condense
in the disk.  In the pulsar-planet formation scenario of Lin, Woosley
\& Bodenheimer (1991), some objects as large as $\sim 10^{21}$ gm formed
much earlier, after $\sim 10^3$ yrs, but $\gsim 10^6$ yrs were needed to form
planets.  Planets might be necessary to deflect a significant number
of objects toward the neutron star (e.g., van Buren 1981; Tremaine \& 
$\dot{\rm Z}$ytkow 1986),
so a delayed turn-on as Colgate \& Leonard propose seems feasible.   

Although the gross energy requirements of halo GRBs can easily be met
in an accretion-powered model, these models have two potential problems:
(1) producing hard non-thermal gamma ray spectra without over-producing
thermal X-rays; and (2) producing a large enough rate of abrupt planetesimal 
captures (eq.~[\ref{eq-repe}]).

Some form of magnetically-mediated gamma-ray production is usually invoked
to solve the first problem (e.g., Colgate 1992).  That is, the accreting
matter distorts the magnetic field, which---via nonlinear Alfv\'en wave 
damping, tearing instabilities, or
some other mechanism---accelerates charged particles to produce gamma rays.  
Note, however, that the {\it total} exterior magnetic energy
of a neutron star with $B_{\rm dipole} = 10^{12} \ B_{12}$ G is less than
the energy of a single GRB in the galactic halo by 
\begin{equation}
\label{eq-ratcol}
 {E_B\over E_{\scriptstyle GRB}} = 0.2 \ B_{12}^2 \ D_{100}^{-2} \ 
\left( { \Delta \Omega \over  4 \pi}\right)^{-1} \ .
\end{equation}
It is difficult to understand how such a relatively large accretion 
energy (eq.~[\ref{eq-eng}]) 
could be channeled through such a weak field without enormous 
inefficiency, resulting 
in the over-production of thermal X-rays.   This problem is discussed
in more detail in \S 7.3.1 of TD95.\footnote{TD95 focus on
accretion-powered models of the 1979 March 5th event's hard initial
spike, but the spectral requirements are similar (FKL).} 

A strong magnetic field is also generally invoked to increase the 
capture cross-section of a neutron star, thereby enhancing the
putative GRB rate (Colgate \& Petchek 1981; van Buren 1981).  
A crude estimate of the radius $R_c$ 
at which magnetic capture will occur in a single orbital pass is given 
by the condition that the kinetic energy of mass $M_p$ at $R_c$ be 
comparable to the magnetic
dipole field energy external to $R_c$. \  This yields 
\begin{equation}
\label{eq-caprad}
 {R_c\over R_\star} = \left({ R_\star^4 \, B_{\rm dipole}^2 \over 
6 \, G M_\star M_p}\right)^{1/2} 
=1.7 \ \ B_{12} \ D_{100}^{-1} \ \epsilon_\gamma^{1/2} 
\left( { \phi_b \over  20^\circ}\right)^{-1} 
\end{equation}
for $R_\star = 10$ km and $M_\star = 1.4 \, M_\odot$.  Thus the single-orbit
capture cross-section is barely enhanced over the geometrical cross-section
of the star for pulsar-strength fields and halo locations.
Multi-orbit magnetic captures also occur (van Buren 1981); but as
shown by Woosley \& Herant (1996), very oblique captures 
produce extended soft X-ray events.

For local galactic disk bursters ($D_{100}\sim 10^{-3}$), the context 
in which planetesimal accretion was 
originally proposed, neither of these problems (as roughly illustrated by
the small numbers in eq.~[\ref{eq-ratcol}] and [\ref{eq-caprad}]) are 
serious.  However, for halo GRBs, dramatic beaming ($\Delta \Omega \ll 1$)
or magnetic fields much stronger than $10^{12}$ G, or both,
seem theoretically desirable.

Thus we suggest that {\it planetesimal accretion onto magnetars} 
may power halo GRBs.  Some subset of the rapidly-rotating proto-neutron
stars which become magnetars (according to the conjecture described in DT96)
may shed excess angular momentum into low-mass disks (eq.~[\ref{eq-tplanet}]),
or capture disk material as they move through clumpy supernova ejecta
(Woosley \& Herant 1996).
The exterior magnetic energy of these stars is greater than 
the energy of a halo GRB by $\sim 2\times 10^5 \ B_{15}^2 \ D_{100}^{-2}
\ (\Delta \Omega/4\pi)^{-1}$, potentially 
catalyzing substantial non-thermal emissions.  The single-orbit 
capture cross-section for planetesimals is enhanced over the geometrical 
cross section by a similar factor.  

The HBM could apply in a magnetar-accretion model 
because the mechanisms for stellar recoils and for aligned beaming 
described in \S 2 would hold.  Alternatively, a time
delay for planetesimal formation (Colgate \& Leonard 1994) may operate (LD92).
This physical model for halo GRBs is viable even if all 
magnetically-powered activity terminates after the SGR phase, as 
long as a magnetar retains a substantial fraction of its initial dipole field,
anchored in the stably-stratified core, for a time 
$\gsim 10^8 \, V_3^{-1} \, D_{100}$ years. 

\section{ Model Extensions and Refinements}
\label{sec-hbm-ext}

\subsection{Intensity Variations Across the Beam}

So far we have assumed that the gamma ray intensity is uniform within 
a beaming cone of half opening-angle 
$\phi_b$. \ In this section we will study the implications of a gamma-ray
intensity which varies as a function of the angle from the beam axis. 

We will use the relativistic beam model of Mao \& Yi (1994).
In this model, gamma-ray emitting material  is expanding 
with Lorentz factor $\gamma\equiv (1-v^2/c^2)^{-1/2}$
into a cone with half opening angle $\Delta\theta$. \ Emissions
are assumed to be isotropic in the rest-frame of the emitting material,
which means that they are beamed to within an 
opening angle $\sim \gamma^{-1}$ about the material velocity vector in
the observer's frame. 
Integrating over the material beam and over photon emission angles, 
Mao \& Yi (1994) derived the net observable photon 
intensity as a function of angle $\theta$ away from the beam axis:
\begin{equation}
\label{intensity}
I(\theta) = \Lambda \, \int_0^{2\pi} \,  d\phi^\prime \ \int_0^{\Delta \theta}
\, \hbox{sin} \theta^\prime \ d\theta^\prime \ \chi \, 
\left[ \gamma (1-[v\chi /c]) \right]^{-(2+\delta)} .
\end{equation}
In this equation, $\chi \equiv \hbox{cos}\theta^\prime \ \hbox{cos}\theta -
\hbox{sin}\theta^\prime \ \hbox{sin}\phi^\prime \ 
\hbox{sin}\theta$; \ $\delta$ is the photon
spectral index (we will take $\delta = 2$); and $\Lambda$ is a normalization
factor, which we will define such that $I(0) = 1$ for all values of 
$\Delta \theta$ and $\gamma$. 

This idealized physical model provides a useful two-parmeter family of  
beams with centrally-peaked brightness.  
Yi and Mao (1994) used this model to show that relativistic beaming 
effects in galactic halo GRB models can produce anti-correlations 
between the brightness and duration of observed bursts, 
mimicking cosmological time dilation.\footnote{When
bursters are unbound from the Galaxy, as in the HBM, 
more distant sources are systematically older.  Any trend of spectral
softening or increasing burst duration 
{\it with age} could then mimic cosmological time dilation.} 
Such correlations were found by Norris et al.~(1993) in the
BATSE data, but not by Mitrofanov et al.~(1996).  

In Figure \ref{fig-beam-int} we plot $I(\theta)$, as given by equation
(\ref{intensity}). \  We have kept $\Delta \theta = 20^\circ$ in this 
plot, which ensures that the beam half-width at half-maximum
stays close to $20^{\circ}$. \ (This is likely to fit 
BATSE and PVO data best; see eq.~[3].)  Figure 7 shows that, for small
 $\gamma$,  bursters become 
potentially detectable at large $\theta$ ($> 20^{\circ}$), 
although only at low intensity, i.e., only  
within a diminished sampling depth by BATSE. \  
The $\phi_b = 20^\circ$ uniform beam case, considered in \S 3 and \S 4 
(solid line in Figure 7), is equivalent to the limit $\gamma \to \infty$
on this plot.

What values of $\gamma$ are plausible for a halo GRB model? \ 
A pure photon/pair fireball (presumably channeled by the
magnetic field into a restricted solid angle) with intrinsic luminosity 
appropriate for a galactic halo GRB would have 
$\gamma\sim 10$ or slightly less (\cite{ps93}).
This is comparable to the value used by Yi and Mao (1994)
in their halo ``time dilation" study.
The Lorentz factor could be moderately reduced, say to $\gamma \sim 3$,
by a plausible contamination of baryons.\footnote{ Of course, fireballs
produce (blueshifted) thermal-spectrum photons, which is {\it not} what
GRBs look like; however this gives at least a first estimate of what
the {\it hydrodynamically-driven} Lorentz factor would be.  Electrodynamic
acceleration (e.g., driven by magnetic reconnection or Alfv\'en waves)
might produce larger $\gamma$, and/or different beam patterns.} 

To study the implications of these $I(\theta)$ models, 
we assume as before that the axi of beamed 
GRB emissions are aligned with peculiar recoil velocities 
${\bf V_r}$ in a galactic population of bursting stars,  
and we do new Monte Carlo calculations of the observable GRB distributions. 
Specifically, for all the $I(\theta)$ functions shown in Figure 7, we 
repeated the log ${\cal N}$---log $P$ model fitting against 
PVO and 3B data (as shown in Figure 4 for the uniform beam case),  
finding the best-fit peak photon luminosity, now defined at
the beam center ($\theta = 0$).  We then used this best-fit 
value to calculate the cumulative angular statistics as a function of peak 
photon flux, similar to Figure \ref{fig-ang-hbm-3b}.
We show these results in Figure \ref{fig-ang-diff}.  For clarity, 
we have plotted the {\it difference} of the angular 
statistics $\cost$, $\sinb$ and $\coss$ from the values predicted 
in the uniform beam case (i.e., the $\phi_b =20^\circ$ HBM values, as
plotted in Figure \ref{fig-ang-hbm-3b}), for 
various intinsic beaming patterns $I(\theta)$. \

Figure 8 shows that, as long as the 
half-width at half-maximum of the gamma-ray beam is held constant, 
observable results for various beam intensity patterns $I(\theta)$ deviate 
little from the uniform beam case.  (Note that the vertical axis scale  
in Figure \ref{fig-ang-diff}  is expanded by a factor 
$\sim 10$ relative to that of Figure \ref{fig-ang-hbm-3b}.) 
This is our main conclusion. In fact,
all of the $I(\theta)$ beaming patterns considered in Figure 7 and Figure 8
fit the present 3B data with better than 3--$\sigma$ confidence at all
observed peak photon flux levels, as can be seen by comparing Figure 8
with the BATSE data of Figure \ref{fig-ang-hbm-3b}. \
Thus a uniform beam is {\it not} required in the HBM, as long
as the variation within the beam is not too large.

\subsection{Burst Luminosity Functions}

We now relax the ``standard candle" assumption (point [4] in the second
paragraph of \S 2.1). We will consider GRB luminosity functions of the form
\begin{equation}
\label{eq-lumfunc}
\Phi(L) \propto L^{-\beta} \qquad\quad \hbox{for} \qquad  
L_{\rm min} < L < L_{\rm max}.
\end{equation}
Constraints on GRB luminosity functions have been previously inferred 
from BATSE and PVO data by 
Ulmer, Wijers \& Fenimore (1995, hereafter UWF),  Hakkila et al.~(1995),
and by other workers cited therein.
In these studies, the bursters' spatial density distribution
in the galactic halo was assumed to be 
\begin{equation}
\label{eq-galdens}
 \rho(r) = { \rho_0 \over [1+(r/R_c)^{\alpha}] } \ .
\end{equation}
The intensity distribution of GRBs (log ${\cal N}$ -- log $P$ curve) 
in the HBM resembles that of the $\alpha = 2$ case of 
this model, although the angular distribution of 
observable bursters as a function of $r$ is very different.

These previous studies concluded that the acceptable 
dynamic range ($L_{\rm max}/ L_{\rm min}$) for GRBs {\it observed}
by BATSE is quite limited.  In particular, various studies agree 
that 90\% of detected bursts must have peak luminosities
within a factor of $\sim 10$, for a wide range of indices $\alpha$ and
$\beta$.  \ The basic reason for this is that
the turnover of the observed log ${\cal N}$ -- log $P$ curve (Fig.~4) 
from a slope 
$d \, \hbox{log} {\cal N} (\hbox{$>$}L)/ d \, \hbox{log} L = -3/2$ 
(the ``Euclidean core") to a shallower slope at fainter fluxes 
happens rather sharply.
A GRB luminosity function that spanned much more than one order of magnitude
would smooth out this turnover to an extent inconsistent with
observations. 

In the $\alpha = 2$ case (relevant to the HBM) UWF found, furthermore,
that the dynamic range of the luminosity function, $L_{\rm max}/ L_{\rm min}$, 
can only be as large as $\sim 10$, and still fit observations,
if the power-law index lies in the range $\beta = 2-3$. \  (See Fig.~2 
in UWF).  Other values of $\beta$ would require significantly more restricted
dynamic ranges.  This result can be understood analytically. 
If we assume $L_{\rm max} / L_0 = L_0 / L_{\rm min} \equiv \kappa$
where $L_0$ is the best standard-candle (mono-luminosity) fit to the data, 
then simple integrals show that the average luminosity satisfies 
$\bar L /L_0 \approx 1$ for $2< \kappa < 20$ only when $\beta \sim 2-3$. \ 
$\bar L/L_0$ is very different from unity when $\kappa> 3$ 
for all other values of $\beta$. \ 
Since luminosity functions with $\bar L/L_0 \sim 1$ give the best 
fits to observations, the dynamic range, given by 
$\kappa^2 = L_{\rm max}/L_{\rm min}$, can only be large for
$2<\beta<3$.   More general analytic constraints on GRB luminosity functions 
have been derived by Horak et al.~(1996).

These above-quoted results were all obtained from fits to the 
the observed log ${\cal N}$---log $P$ curve only. 
Next we consider constraints on the GRB luminosity function 
from the observed log ${\cal N}$---log $P$ distribution 
{\it combined} with the  observed angular 
distributions.  These combined constraints were previously considered by 
Hakkila et al.~(1995)
for a galactic halo of bursters with density given by 
eqn.~(\ref{eq-galdens}) and no preferred beaming directions.
Here we focus on the very different
case of the HBM.  (Of course, the GRB angular distribution cannot 
constrain the luminosity function of {\it cosmological} GRBs,
since the bursters are distributed essentially isotropically around Earth
at all intensity levels.)

Intuitively, luminosity functions could either wash out 
or enhance the angular anisotropies predicted by HBM 
for bright subsets of GRBs (Fig.~6).  If the luminosity function 
of bursts is a positive (negative) power-law, then 
at a given observed flux level, a greater (smaller) number of more
distant bursters are counted than in the mono-luminosity case.
Consequently, the degree of anisotropy is reduced (enhanced) 
since the HVNS distribution appears more and 
more isotropic at greater distances from the Earth. 

To illustrate this, we consider two cases: 
$(\beta, \kappa) = (-3, 2), \, (3, 4.4)$. \
Both models give acceptable fits to the log ${\cal N}$---log $P$
distribution of the combined PVO and BATSE data, but 
the allowed dynamic range $L_{\rm max}/L_{\rm min} = \kappa^2$
is much smaller in the $\beta = -3$ case, for reasons discussed above.
Figure \ref{fig-ang-hbm-lum-3b}
shows corresponding angular distribution statistics (dashed and
dotted curves) compared to the mono-luminosity case (thick solid curve). 
It is clear that the luminosity function with positive 
slope ($-\beta = 3$) affords a better fit 
to the angular position data than the standard--candle model does. 
For $\beta = 3$, on the other hand, 
the angular statistics deviate from BATSE by $\sim 3$--$\sigma$ 
at the faint end. (Plotted error bars are 3--$\sigma$.)

We conclude that only a limited spread of intinsic luminosities
among GRBs observed by BATSE and PVO are admissible in the HBM, as in 
other galactic halo models.  Depending on the exact form of the 
luminosity function, this could either improve or worsen the model's fit
to the BATSE angular distribution.  However, in either case the effect
is limited because the dynamic range is small.  Equivalently, one could 
say that in a fit to the {\it combined} GRB brightness and angular position 
data, angular information increases (decreases) 
the statistically--allowed dynamic range of
the luminosity function when $\beta <0$ ($\beta >0$), but this 
dynamic range is less than $\sim 10$ in any case.

Note that we have considered only  
one possible generalization of the basic (standard) HBM here: 
a power-law luminosity function that is uniform among bursters. 
Other conceivable generalizations, such as 
luminosity functions that vary with burster age, or luminosities correlated
with $\phi_b$ or $V_r$, would produce different results.

\subsection{Weakly-bound Bursters in the HBM}

Next we consider relaxing the assumption that all bursters are born
with $V_r = 1000$ km s$^{-1}$ (point [2] in the second paragraph of
\S 2.1).  If some fraction of bursters are born with velocities
less than the local galactic escape velocity, which is about 
$V_{esc}\sim 600$ km s$^{-1}$ at typical birthsites in the disk,
these stars will orbit the galaxy.  Podsiadlowski, Rees and Ruderman
(1995) have emphasized the important point that these {\it weakly-bound 
bursters do not return to the small galactic radii at which they were born} 
because the gravitational potential in the
halo is probably markedly non-spherical. Large deviations from
spherical symmetry are predicted by almost all models of galactic halo
formation (e.g., Dubinski 1992 and references therein).  
Thus weakly-bound bursters (i.e., bursters born with velocities narrowly
below the local galactic escape velocity) will tend to 
stay in the outer halo, following wide-ranging, non-periodic orbits.
After one or more orbits, their magnetic axi orientations are 
randomized. If these weakly-bound bursters continue to emit (beamed) bursts, 
their contribution to observed GRBs must be added to the bursts 
from the escaping ones (i.e., the unbound ``standard HBM" sources, studied in
\S 3 and \S 4).  This could, under some conditions that we will discuss below, 
produce a net angular distribution of GRBs that is 
more isotropic than that of the standard HBM, especially at the faint end.
However the basic HBM mechanism for producing the ``Euclidean core" in 
the GRB brightness distribution, as observed by PVO, may still
apply. 

To quantify this discussion, assume that a fraction $f = 10^{-1} f_{-1}$
of all bursters are born with velocities below the local galactic escape
velocity, and the remainder have $V_r = 1000 \, V_3$ km s$^{-1}$ as before.
This is a crude way to estimate the effects of a realistic (Gaussian) 
velocity distribution. If the total birthrate of GRB sources in the galaxy 
is $\Gamma = 10^{-3} \, \Gamma_3$ yr$^{-1}$ (eqn.~[\ref{eq-brate}]), 
then the number of bound bursters presently orbiting in the halo is 
\begin{equation}
\label{eq-numbound}
{\cal N}_{\rm bound} = \Gamma \, f \, \tau_{\rm gal} = 1 \times 10^6 \ \Gamma_3
\, f_{-1} ,
\end{equation}
where $\tau_{\rm gal} \sim 10^{10}$ yr is the age of the galaxy
(i.e., the period during which bursters have been forming).  The
relative abundance of bound bursters compared to unbound ones 
(eqn.~[\ref{eq-nhalo}]) within  the BATSE sampling depth of 
$D= 100 \, D_{100}$ kpc is then
\begin{equation}
\label{eq-ratiobound}
{{\cal N}_{\rm bound} \over {\cal N}_{\rm unbound}} 
= 10 \ f_{-1} \ D_{100}^{-1} \ V_3 .
\end{equation}
Thus if $> 1\%$ of bursters remain bound to the galaxy ($f > 10^{-2}$),
bound stars will numerically dominate the halo (DT92).
If the bursters continue to emit bursts within beaming cones of 
half angle $\phi_b$, but with orientations randomized by
their orbits, for a bursting lifetime
$\tau_{\rm life} = 10^{10} \, \tau_{10}$ years, then the fraction of
all {\it bound} bursters that are observable at Earth is
\begin{equation}
\label{eq-fracobs}
(1 - \hbox{cos} \, \phi_b) \ {\tau_{\rm life}\over \tau_{\rm gal}} = 0.06
\ \tau_{10} \, \left({\phi_b \over 20^{\scriptscriptstyle o}}\right)^2 .
\end{equation}
Multiplying equation (\ref{eq-ratiobound}) and equation (\ref{eq-fracobs})
yields the relative number of {\it observable} bound bursters compared 
to unbound ones: 
\begin{equation}
\label{eq-psibound}
\Psi_b = 0.6 \ f_{-1} \ \tau_{10} \ D_{100}^{-1} \ V_3 \  
\left({\phi_b \over 20^{\scriptscriptstyle o}}\right)^2 .
\end{equation}
It is interesting that this can be of order unity.

A fraction  $\Psi_b / (1 + \Psi_b)$ of all observed GRBs will 
come from bound (orbiting) bursters. How anisotropic will be the sky
distribution of these ``bound bursts"? \  This depends upon several 
uncertain factors, especially the burster recoil velocity distribution
and the form of the galactic halo potential.  Here we limit ourselves 
to a few general comments, based upon the results of PRR.

If many bursters are born with $V_r \lsim 500$ km s$^{-1}$, 
significantly below the escape velocity from the center of the galaxy, 
then the bound burst distribution will  
be {\it very} anisotropic (cf.~Fig.~2a and 2b of PRR).  If, however, most
bound bursters aquire velocities only narrowly below the local escape
velocity, then the bound burster distribution could be nearly isotropic,
as shown in Fig.~5b of PRR.  This figure shows the burster distribution for 
kick velocities uniformly distributed on the range
$700 \le V_r \le 900$ km s$^{-1}$, and a realistic, nonspherical galactic
potential. 

In the magnetar model of GRBs, burster recoils are imparted by a complicated
series of magnetically-induced neutrino impulses (\S 2.2), with durations 
comparable to the coherence time of magnetic structures near and below
the neutrinosphere (DT92; TD93).  Because these impulses are additive,
the central limit theorem implies that a Gaussian with mean $\overline V$ and
variance $\sigma_{\scriptscriptstyle V}$ is probably a good first 
approximation to the $V_r$ distribution among bursters.  The precise values 
of $\overline V$ and $\sigma_{\scriptscriptstyle V}$
are uncertain, although SGR data suggest $\overline V \gsim 1000$ km s$^{-1}$
(Appendix A).  Thus the $V_r$ distribution for {\it bound} bursters is
probably the low--$V_r$ tail of a Gaussian.  If 
$\sigma_{\scriptscriptstyle V} \lsim 200$--300 km s$^{-1}$, 
then the velocity distribution of
bound bursters is sharply peaked just below the escape velocity, and 
conditions for a nearly isotropic sky distribution of bound bursts could
be met.  The distribution of bound bursters might still show mild
``hot spots" in some presently--unknown directions associated with 
distortions in the outer galactic halo potential (PRR), but these
anisotropies might not be apparent from the statistical tests used 
in this paper. [More general statistics (e.g., Briggs et al.~1996;
Tegmark et al.~1996) would be useful for constraining this effect.]

Assuming that $\sigma_{\scriptscriptstyle V} \lsim 200$--300 km s$^{-1}$ 
and that other 
conditions for a nearly--isotropic bound burst distribution are met, then 
the total model galactocentric quadrupole moment would be
\begin{equation} 
\label{eq-modsinb} 
\langle \hbox{cos}^2 \Theta\rangle - 1/3 = 
(\Psi_b + 1)^{-1} \ \big[\langle \hbox{cos}^2 \Theta\rangle_{\rm HBM} - 
(1/3) (1-\Psi_b) \big] ,
\end{equation}
where $\coss_{\rm HBM}$ is the moment in the ``standard HBM", as shown
in Figure (6), and we have assumed that the bound burster distribution 
has $\coss = 1/3$.  Thus for $\Psi_b = 1$, the deviation from
isotropy would be reduced by (1/2) compared to the standard HBM. \ Similar
results might apply for $\cost$ and $\sinb$; however it is also possible 
that $\sigma_{\scriptscriptstyle V} \gsim 300$ km s$^{-1}$ 
and that the addition of bound stars 
makes the model fits to BATSE observations {\it worse}.  We will study 
this issue in a future paper.   
 
Note that the mean time between bursts from a given star is {\it longer}
by a factor $(1+\Psi_b)$ when a population of bound bursters is present:
\begin{equation}
\label{eq-boundrepe}
\taugrb = 110 \ (1 + \Psi_b) \ \Gamma_3 \ V_3^{-1} \ D_{100} \ \hbox{yrs}
\end{equation}
(cf.~eq.~[\ref{eq-repe}]).  Despite this fact, the energy required 
for each burster is significantly greater, because of the long bursting 
lifetime:
\begin{equation}
\label{eq-boundteng}
E_{tot} = 5\times 10^{48} \ \tau_{10} \ D_{100} \ \Gamma_3^{-1} \ V_3 
\ (1+\Psi_b)^{-1} 
\ \left({\phi_b^2 \over 20^\circ } \right)^2 \
\hbox{erg}. 
\end{equation}
This still might be supplied by magnetic energy (cf.~eq.~[\ref{eq-beng}]).
Alternatively, if the interior flux density satisfies $B< 3\times10^{15}$ G 
initially, then from eq.~(\ref{eq-bbeng}) and eq.~(\ref{eq-boundteng}) we infer
that magnetically-powered bursting would have to terminate after an active 
lifetime no longer than 
\begin{equation}
\label{eq-boundlife}
\tau_{\rm life} = 10^9 \ B_{15} \  \Gamma_3 \  V_3^{-1} \ D_{100}^{-1} \ 
(1+\Psi_b) \ \left({\phi_b^2 \over 20^\circ } \right)^2 \
\hbox{yrs}.
\end{equation}
This corresponds to $\tau_{10} = 0.1$ in eq.~(\ref{eq-psibound}),
suggesting that magnetically--powered, orbiting bursters might cease
GRB activity too early to make a substantial contribution to the
observable burst distribution ($\Psi_b \ll 1$).  However, 
accretion--powered bursters
(\S 5.2) would not be subject to this bursting lifetime limit.

\section{Concluding Discussion}
\label{sec-conclu}
\subsection{Summary}
 In this paper we have considered models for GRBs from 
high-velocity neutron stars in the galactic halo.  In particular, 
we compared GRB observations with a model having the following properties:
[1] bursters are born at positions distributed like young Pop.~I stars
in the galactic disk; [2] with randomly directed recoils
$V_r = 1000 \, V_3$ km s$^{-1}$; [3] they emit GRBs at a constant rate, 
with [4] constant luminosity, and [5] the gamma ray emission 
is beamed parallel and anti-parallel to ${\bf V_r}$, 
within an angular radius $\phi_b$. 

This ``standard halo beaming model" (S-HBM) adequately fits 
all published data on the distributions of GRBs 
from the BATSE and PVO experiments, when  
$12^\circ \, V_3^{-1} < \phi_b < 30^\circ$.
\ Models for GRBs from sources at cosmological distances fit present BATSE 
observations {\it better} than the S--HBM does, but not by a 
statistically-compelling margin.  These fits to the data, for 
$V_r = 1000$ km s$^{-1}$ and $\phi_b = 20^\circ$,  are
shown in Figure 4 and Figure 6.  Note that the fit 
improves for $V_r > 1000$ km s$^{-1}$ (as found in most SGRs; 
Appendix A).

An important feature of the HBM is that
{\it almost all bursters born in the Andromeda Galaxy (M31) or
other nearby external galaxies are 
unobservable at Earth because of misdirected beaming} 
(LDT; Li, Fenimore and Liang 1996).  Only when the sampling depth 
of the instrument reaches M31 can the effects of M31 possibly become
evident. This means that the often-quoted upper bounds on the BATSE 
sampling depth based on considerations of M31 (e.g., LD92; \ Hakkila et 
al.~1994) do not apply to the HBM. \
Bounds on halo models based on the nondetection
of X-ray bursts from M31 (Li \& Liang 1992) or in 
samples of nearby external galaxies 
also do not constrain the HBM, as long as the X-rays 
associated with GRBs are beamed in a similar way as the gamma rays.  

We have also considered three refinements to the S--HBM, 
relaxing the assumptions [2], [4] and [5] in the
above list:

{\it Intensity variations across the beam.}  
We analyzed models with centrally-peaked gamma-ray emission beams; 
i.e., with intensity declining as a function of angular separation from 
the beam axis, as in relativistic emission models.  We found that 
the observable model results are not very sensitive
to the details of the beam intensity distribution.
Adequate fits to the data were obtained as long as the
angular radius at which the intensity falls to half its central
value is $\sim 20^\circ$, and the angle at which it drops by another
factor $\sim 2$ is not many times larger. 

{\it Burst luminosity functions.}   
We explored the consequences of a spread in GRB intrinsic brightnesses,
considering luminosity functions of the form 
$\Phi(L) \propto L^{-\beta}$  for   $L_{\rm min} < L < L_{\rm max}$. \
We found that, in the HBM as in other galactic halo models, only a limited 
dynamic range $L_{\rm max}/L_{\rm min} \le 10$ or 20 is admissable
among GRBs observed by BATSE and PVO. 
Depending on whether the power-law index satisfies 
$\beta<0$ or $\beta >0$, the spread in luminosities 
either improves or worsens the model's fit
to the BATSE angular distribution.  However, in either case the effect
is small because the dynamic range is limited.

{\it Models with a subpopulation of bound bursters.} 
In the S--HBM, all bursters are born with recoil velocities 
$V_r= 1000 \, V_3$ km s$^{-1}$. \  Realistically, one 
expects bursters to aquire a distribution of $V_r$ values.  \ 
The low--$V_r$ tail in this distribution could 
correspond to bursters that remain weakly-bound to the galaxy.  
Podsiadlowski, Rees \& Ruderman (1995) have shown 
that these bound stars follow non-periodic orbits
in the non-spherical galactic halo potential, generally not returning
to the small galactic radii where they were born. 
After one or more orbits, their magnetic pole (beaming axis) 
orientations are randomized.
If they continue to emit (beamed) GRBs at late ages, then 
the bound and escaping bursters might produce comparable 
numbers of observable GRBs. 
(The exact ratio, eq.~[\ref{eq-psibound}], depends 
upon several uncertain model parameters.)  
The observable distribution of GRBs would then be 
an appropriately weighted sum 
of the S--HBM burst distribution, describing the unbound 
bursters (Fig.~4 and Fig.~6), and the bound burster 
distribution studied by PRR.  \  The resultant model angular distribution
might be significantly more isotropic than that of the standard HBM, if
the orbiting bursters are only weakly bound to the galaxy (see \S 6.2).
This plausible refinement of the standard HBM merits further study. 

\subsection{Conclusion: Cosmological or Galactic GRBs?}

Our main conclusion is as follows.
The models studied in this paper give acceptable fits to observations 
as long as certain special conditions (e.g., uniform bursting rate, 
$12^\circ \, V_3^{-1} < \phi_b < 30^\circ$) are satisfied. 
These conditions, although possible and arguably plausible, have no 
clear and compelling theoretical justification.
Other special conditions of varying plausibility 
are required in other halo models (e.g., LD92; see Hartmann et al.~1994
for a comprehensive review of halo models).
The cosmological hypothesis, on the other hand, fits present observations 
{\it generically,} and 
thus is much more attractive in this sense.  Moreover present data 
are generally better fitted by the cosmological models than by the 
S--HBM, as shown by Fig.~6.

Based upon this powerful argument, we believe cosmological
GRB models to be more promising than galactic halo models, given the
evidence available at the time of writing.
However, some caution is appropriate because there exists no
compelling theoretical understanding of GRBs at either distance scale.
The study of GRBs has taken many unexpected twists and
turns in the past:  only five years ago there was a wide consensus
favoring neutron stars at distances $\lsim 10^2$ pc.

If the cosmological hypothesis is correct, then continued BATSE data
collection should be able to rule out the standard HBM using the 
test of Fig.~6.  A doubling of the data base beyond the 3B catalog 
might rule out the S-HBM with $> 3$--$\sigma$ confidence. 

 A plausible refinement of the standard HBM that might be more difficult
to falsify includes some 
bursters that are weakly-bound to the galaxy (\S 6.3).  
Such a model can be {\it jointly} constrained by limits on
GRBs---or associated X-ray bursts---from nearby {\it external} galaxies
(e.g., Li \& Liang 1992),
which limit the bound burster population, and BATSE angular position 
observations, which more strongly limit the unbound ones.
We will quantify these constraints in a future study.

If GRB {\it do} come from the galactic halo, then 
deviations from isotropy on the sky will eventually
manifest themselves. In any case,  
{\it continued operation of BATSE is essential for making
these tests.} Checking BATSE with an independent experiment is 
also desirable.  It is important to verify that systematic errors
do not wash out the very low-level anisotropies predicted by 
halo models.  

GRB repetitions, if verified, could lend 
credence to halo models; however the predicted mean time 
interval between repeat bursts from old SGRs in the halo 
is $\taugrb \sim 100$ years, with some dependence on uncertain
model parameters (eq.~[\ref{eq-repe}] and eq.~[\ref{eq-boundrepe}]), 
so repeat events may be difficult to detect unless they are strongly 
clustered in time.\footnote{Magnetically-powered 
GRBs would tend to cluster in time, like
stellar flares, so incidents of GRB repetition should 
eventually be observed if the bursters are isolated magnetars. 
This could be an important
observational diagnostic, because most cosmological GRB models
do not allow burst repetitions.}   
\footnote{ Many bursts in the BATSE catalog show distinct episodes 
of bursting activity  separated by $\sim 10$--100 s intervals 
containing no detected emissions (Lingenfelter, Wang \& Higdon 1994).  
It would be interesting to search for such recurrences on somewhat 
longer timescales, $\sim 10^2$ s to $10^4$ s, in order to place bounds on 
how long the time window for recurrence lasts.  Unfortunately, BATSE is 
biased against recording data during a readout period lasting from
$\sim 100$ s to $\sim 90$ minutes 
following any burst, when it it downloading data.  A large fraction
of bursters could produce lower-intensity recurrences during this period 
without any detections.} 
 
The durations of BATSE bursts were found to be anti-correlated 
with their intensities by Norris et al.~(1993) [but see 
Mitrofanov et al.~1996].   This could be due to  
cosmological time dilation.  However, this correlation, if verified,
does not give strong evidence {\it against} the HBM. \  In the HBM 
more distant burst sources are systematically older, so 
any significant trend of increasing burst duration or spectral 
softening with age could mimic cosmological time dilation.

What are the strongest arguments {\it in favor} of galactic halo 
GRB sources? \
There is strong empirical evidence that a large number of
old soft gamma repeaters exist in the extended galactic halo (DT92). 
We estimate the number of aged SGRs within 100 kpc to be between 
$\sim 10^4$ and $\sim 10^7$, depending on values of the scaling parameters 
in equations (\ref{eq-nhalo}) and (\ref{eq-ratiobound}). 
These halo stars have many 
properties required of classic GRB sources, as first pointed out by DT92.  \ 
They are clearly distinct from radio pulsars (footnote 12 and Appendix A),
thus avoiding the problem of over-concentration of bursters toward 
the galactic disk which plagues many theories based on neutron stars. 
Moreover, of all known astrophysical objects, SGRs 
have produced the emissions which most closely resemble GRBs.  They 
have produced at least one burst of photons with a classic GRB-like
spectrum (on March 5, 1979; FKL); and when they are young 
(ages $\lsim 10^4$ yrs), they emit large numbers of softer-spectrum 
bursts with energies $\sim 10^{41}$ ergs, comparable to  
the energies of GRBs in the halo.  

We have briefly outlined some candidate physical 
models for these bursters (\S 2.3, \S 2.4, and \S 5).  
One promising class of models involves neutron stars with
ultra-strong magnetic fields $B > 10^{14}$ G, much stronger than the
fields known to exist in ordinary radiopulsar magnetospheres.
Such strong magnetic fields could provide the energy for 
flare-like halo GRBs (\S 5.1), or they could 
mediate the conversion of accretion energy into gamma rays (\S 5.2).  
Fields this strong could also drive crust fracture events in young neutron
stars, producing SGR activity (TD95).  [Note that the distributions 
of SGR burst energies and waiting times, and their time-correlations, 
are consistent with crust fractures; Cheng et al.~1996.]  Finally, strong 
magnetic fields can drive neutrino emission anisotropies in newborn 
neutron stars, imparting recoils that propell 
the stars into the galactic halo (\S 2.4).
Note that all of these mechanisms could conceivably operate in a non-HBM 
context, i.e., a halo GRB model in which the bursts are unbeamed or beamed in 
random directions (see footnote 14).  

If aged SGRs do {\it not} emit classic GRBs,
then it is still likely that these stars populate the
galactic halo (DT92).  If they are isolated magnetars, then they will 
rotate very slowly after leaving the galactic disk.  They will cool rapidly 
once their magnetic activity ceases, becoming dark halo relics that are 
difficult to detect.  

Several promising models of {\it cosmological} GRBs also involve
ultra-strong magnetic fields (Thompson 1994 and references
therein).  As a general result, 
gravity--powered MHD flows in compact neutron objects can 
create fields of strength $ B^2/ 8 \pi \sim \epsilon \, G M^2/ R^4$,
where $\epsilon$ is the efficiency.   This implies
\begin{equation}
\label{eq-Bmax}
B\sim 10^{18} \ \left({\epsilon\over 0.1}\right)^{1/2} \  
\ \left({M\over M_\odot}\right) \  
\left({R\over10 \, \hbox{km} }\right)^{-2}  \ \ \hbox{G}.  
\end{equation}
Although strong by most standards,
this is vastly below the fundamental limit of $\sim 10^{49}$--$10^{53}$ G 
at which magnetic fields become unstable to the vacuum creation
of GUT--scale or Planck--scale monopoles. 

Thus, the $\sim 10^{12}$ G
dipole fields of radiopulsars are {\it a million times weaker} than the
strongest possible flux densities attainable, 
however briefly, in neutron star formation and coelescence. 
The mystery of gamma-ray bursts may require us to finally recognize 
and understand this new regime of strong magnetism.

\acknowledgments

We thank Chris Thompson for key contributions to the basic
idea of the HBM, and Edward Fenimore for assistance with fitting
the PVO data, and for many useful discussions.
We also thank an anonymous referee for perceptive remarks which greatly
improved our discussion of bound halo bursters.
This research was supported by the NASA Theoretical Astrophysics
Program, Grant No.~NAG5--2773, and by the Texas Advanced Research
Program, Grant No.~ARP--279. 
HL gratefully acknowledges the support of the Director's Postdoctoral
Fellowship at Los Alamos National Laboratory.

\clearpage
\appendix
%\centerline{\bf APPENDIX}
\section{Observations of Soft Gamma Repeaters and Evidence for Recoils}
We have suggested that old SGRs in our galaxy are the sources of classic GRBs.
This requires that SGRs aquire sufficiently
large recoils to propell them into the galactic halo. 
Here we briefly review the evidence for large recoils in SGRs, and 
discuss the related issue of whether SGRs have normal stellar 
companions.  We also briefly discuss how SGR observations might be
accounted for in the magnetar model (which predicts large recoils, \S 2.4).  
 
{\bf SGR 0526$-$66} \ \ 
The error box for the 1979 March 5 event (\cite{cetal82}) and the 
SGR-like bursts which followed in its aftermath (Golenetskii, Ilyinski 
\& Mazets 1984) lie within 
supernova remnant N49 in the Large Magellanic Cloud.  
A point-like X-ray source\footnote{ That is, a source with an observed 
X-ray image that is statistically consistent with the instrumental 
point-spread function.} 
inside this error box (\cite{rkl94}) may have a thermal spectrum, since 
no hard spectral component has been detected (Murakami 1996; 
Marsden et al.~1996). Its intensity is consistent with thermal
emission from a magnetically-heated neutron star (TD96).  If the SGR
was born at the center of the supernova that produced N49, then the 
displacement of the X-ray source from the remnant center implies a 
{\it transverse} velocity (DT92)
\begin{equation}
\qquad\qquad\qquad\qquad
V_\perp = 1200 \pm 240 \ \ \hbox{km s}^{-1},  
\qquad\qquad\qquad\qquad 
\end{equation}
where we have used
the estimated age of N49, $t_{\rm N49} \approx 5400$ yrs (Vancura
et al.~1992).  The statistically expected 3--D recoil velocity is
larger by a factor $(3/2)^{1/2}$, or $\overline{V} = 1500$ km s$^{-1}$. 

No radio, infrared or optical emissions have been detected from SGR0526--66 
down to faint limits (Dickel et al.~1995; Marsden et al.~1996).
This is not surprizing since a neutron star born with the velocity of
eq.~(A1) could not remain bound in a binary (DT92); and a neutron
star with an 8.0 second rotation period, as clearly shown by the 
1979 March 5 event (Mazets et al.~1979), could not produce
a significant rotation-powered plerion.

\medskip 
{\bf SGR 1900$+$14} \ \ 
This burster has been localized to a position lying just outside the
edge of the galactic supernova remnant G42.8$+$0.6 (Hurley et al.~1994). 
\  Like SGR 0526--66, it has a candidate point-like X-ray counterpart,
just outside the burst error box (Hurley et al.~1996).
Assuming that the SGR was at the center of the supernova that
produced remnant G42.8$+$0.6, one finds (Hurley et al.~1996) 
\begin{equation}
\qquad\qquad\qquad
V_\perp = 1700 
\ \left({D\over 5 \, \hbox{kpc}}\right)
\ \left({t_{\scriptscriptstyle \rm SNR}\over 10^4 \, \hbox{yr}}\right)^{-1}
\ \ \hbox{km s}^{-1},  
\qquad\qquad\qquad 
\end{equation}
where $D$ is the distance from Earth, and $t_{\scriptscriptstyle \rm SNR}$ 
is the age of the supernova remnant (SNR). This implies $\overline{V} \sim 2100$
km s$^{-1}$.

A double infrared source, probably a bound pair of M5 supergiant stars
which is heavily reddened by interstellar dust, has been suggested 
as a counterpart object to SGR 1900$+$14 (Vrba et al.~1996).  This object lies 
$18^{\scriptscriptstyle \prime\prime}$ from the center of the observed X-ray
image (Hurley et al.~1996) and $50^{\scriptscriptstyle \prime\prime}$
from the burst error box (Vrba et al.~1996) in a region of dense
galactic starfields, so the identification is not certain.  
If the infrared binary is a true counterpart to the SGR, then the
association of the SGR with the supernova remnant is almost certainly spurious,
because no known process could impart a recoil of magnitude (A2)
to a pair of M5 supergiant stars.  However, the other SGRs 
are unambiguously associated with $\sim 10^{4}$--year--old SNRs. 
Thus we favor the interpretation that the 
infrared binary is not physically associated, and the SNR is.   
In this case, SGR 1900$+$14 closely resembles SGR 0526--66 in 
all its known (observed) properties. 
 
\medskip
{\bf SGR 1806$-$20} \ \ 
This burster is located within the $\sim 10^4$--year-old SNR
G10.0--0.3 (Kulkarini et al.~1994).  It also has an 
X-ray counterpart (Murakami et al.~1994) which is point-like, but
with a measured {\it non-thermal} spectral component (Sonobe et al.~1994), 
unlike the other SGRs. \  Furthermore, there is a synchrotron radio 
plerion surrounding the SGR (Kulkarni et al.~1994) indicating that the
burster drives an outflow of relativistic particles.
The radio spectral index of the plerion is steeper than that of known
rotation-powered pulsar plerions (Kulkarni et al.~1994).

In the magnetar model, this plerion could be powered by SGR bursts,
as suggested by Kulkarni et al.~(1994) and studied by Tavani (1994) and
Harding (1996). Another possibility is that the radio nebula and 
non-thermal X-rays are powered by quasi-continuous 
low-amplitude Alfv\'en wave emissions from magnetically-induced  
deep crustal fractures (TD95). We favor this second
interpretation, for reasons explained in TD96.  \ The other SGRs do 
not power radio plerions; this might mean that the deep crustal field 
in other SGRs is strong enough to overwhelm lattice 
stresses, driving plastic creep rather than quasi-steady 
low-amplitude seismic activity (TD96). 

Thus in our favored interpretation, SGR 1806--20 has the {\it weakest} 
magnetic flux density in the deep crust of known SGRs. 
Alternatively, it could have the most spatially disordered
magnetic field in the deep crust, which would also drive a higher
rate of crustal fractures. In either case, we would
expect it to also have the lowest recoil velocity according to 
the mechanisms of \S 2.4.  The displacement of SGR 1806--20 from the center
of the SNR is difficult to estimate because of the irregular shape of
the remnant and because of late energy injection into the plerion,
but Kulkarni et al.~(1994) roughly estimate $V_\perp \gsim 500$ km
s$^{-1}$.

This velocity is small enough, and uncertain enough, that the burster could 
have remained bound to a companion if it formed in a tight and massive
binary.  In fact, a massive companion star {\it has} been detected (Kulkarni et 
al.~1995).  It seems to be a  heavily--reddened OB supergiant
(Kerkwijk et al.~1995).  Its physical association with the SGR is
in little doubt since it coincides with the finely-measured VLA position 
of the plerion's central peak (Vasisht, Frail \& Kulkarni 1995). 

Since 1806--20 is the only SGR with a confirmed companion {\it and}
a radio plerion, it is tempting to speculate that the two 
phenomena are related, and that the plerion is accretion--driven.
However, SGR 1806--20 has few properties
in common with known accretion--powered radio nebulae (Vasisht, Frail
\& Kulkarni 1995).  Moreover, the X-ray power is much less than the
required power in relativistic particles (TD96).  Accretion--powered
outflows from neutron stars (e.g., jets) can have only a modest
outflow efficiency, so the ratio would be the other way around
(TD95; TD96). Still, one should keep in mind that 
SGR 1806--20 is much more complex than
the other two SGRs. It is possible that interaction
with, or accretion from, its hot supergiant companion may play some role
in this system, even if the SGR is a magnetar.  More observations are needed. 

If classic GRBs are produced by old halo SGRs, then bursters 
which remain in binaries, like SGR 1806--20, are constrained to
{\it not} produce too many observable classic GRBs as they grow old. 
If they did, the concentration of GRBs toward the galactic plane could 
violate BATSE limits.  In the HBM, this requirement is actually not
very restrictive, because bursters in binaries have effectively random
beaming directions, greatly reducing their observable burst rates
compared to bursters which become unbound.
Furthermore, GRBs from old SGRs in binaries might be suppressed by
the environmental influences of their companions, or 
because values of $V_r$ that are sufficiently small to avoid binary 
disruption only occur---via the mechanism of \S 2.4---in stars with
systematically  
weaker and/or more disordered magnetic fields, which penetrate the
neutron star core less deeply.  Such stars would be expected to
dissipate their smaller resevoirs of magnetic energy relatively quickly, 
showing less magnetic activity (GRBs) late in their lives. 
The fact that the detectable energy loss from SGR 1806--20 
exceeds that of the other SGRs is consistent with this conjecture. 
\clearpage

%\centerline{\bf APPENDIX}
\section{Alfv\'en Wave--Mediated GRBs in Halo Models}

The amplitude $\delta B$ of Alfv\'en
waves required to to carry a luminosity of 
$L_\gamma = 10^{41} \ L_{41} \ (\Delta \Omega_\gamma/ 4\pi)$ erg s$^{-1}$,
as appropriate for halo GRBs (cf.~eq.~[\ref{eq-lum}]), is
\begin{equation}
\label{eq-alfven}
\delta B  \sim \left({4 \pi \, L_\gamma \over c R^2 \, 
\Delta\Omega_A} \right)^{1/2} \sim 7\times10^{9} \ L_{41}^{1/2} \ 
\left({\Delta \Omega_\gamma\over \Delta \Omega_A}\right)^{1/2}
\ \ \hbox{G}
\end{equation}
near the neutron star surface (e.g., Blaes et al.~1989, hereafter BBGM). 
In this equation $\Delta \Omega_\gamma = 2\pi (1- \hbox{cos}\phi_b)$
is the solid angle of GRB emissions, and $\Delta \Omega_A$ is the solid
angle of the neutron star surface through which Alfv\'en waves pass. 
These Alfv\'en waves presumably couple to shear waves in the crust,
which have velocity $V_\mu = (\mu / \rho)^{1/2}$, where the shear modulus
at density $\rho = 10^{11} \, \rho_{11}$ gm cm$^{-3}$ is 
$\mu = 8\times10^{28} \, \rho_{11}^{1.40}$ erg cm$^{-3}$, valid below
neutron drip, $\rho_{11}<4.6$.  \  This value of $\mu(\rho)$ comes
from eq.~(11) of TD95 and a power-law fit to the equilibrium nuclear
parameters in the Baym, Pethick \& Sutherland (1971) equation of state.
For a neutron star with (uniform, vertical) magnetic flux density $B$, 
the coupling 
between Alfv\'en and seismic waves occurs at the density $\rho_x$ where
$B = [4\pi \, \mu(\rho_x) ]^{1/2}$, or 
\begin{equation}
\label{eq-rhox}
\rho_x = 10^{11} \, B_{15}^{1.43} \ \hbox{gm cm}^{-3}.
\end{equation}
Matching the seismic energy flux $F_s = \rho V_\mu^3 \epsilon^2$,
where $\epsilon$ is the wave strain amplitude deep in the crust,
to the surface Alfv\'en energy flux $F_A = c (\delta B)^2 /4\pi$
implies a strain amplitude at the coupling depth of 
\begin{equation}
\label{eq-strain}
\epsilon = {\delta B \over B} \ \left({V_\mu(\rho_x) \over c}\right)^{-1/2} 
\sim 0.3 \ B_{12}^{-1.3} \ L_{41}^{1/2} \ 
\left({\Delta \Omega_\gamma\over \Delta \Omega_A}\right)^{1/2}.
\end{equation}
Since this ignores wave reflection, it should be considered a lower
bound on $\epsilon$. 
This exceeds the dynamical yield (fracturing) strain for
$B \sim 10^{12}$ G. \  
Thus {\it waves with sufficient amplitude to mediate a halo GRB 
would crumble a radiopulsar's crust and thereby dissipate energy,} probably
producing an excess of X-rays.  Only for fields that satisfy
\begin{equation}
\label{eq-waveB}
B > 2\times 10^{13} \ L_{41}^{0.38} \ 
\left({\epsilon_{\rm yield}\over 10^{-2}}\right)^{-0.77} \
\left({\Delta \Omega_\gamma\over \Delta \Omega_A}\right)^{0.38} \ \hbox{G},
\end{equation}
is the vibration amplitude small enough to avoid 
pulverizing the crust.  This condition 
is numerically similar to the magnetic energy requirement of 
equation~(\ref{eq-breq}), but conceptually distinct. 

Note that when equation (\ref{eq-waveB}) is satisfied, there are actually
four distinct physical domains for the propagation of waves in the
crust:  \ {\bf(1)} \ $\rho > \rho_{\rm drip}\approx 4.6\times 10^{11}$
gm cm$^{-3}$. At this depth, the shear waves have amplitude
that varies with density as $\epsilon \propto \rho^{0.18}$, based on
a power-law fit to the Negele \& Vautherin (1973) equilibrium composition. \
{\bf (2)} \ $\rho_x < \rho < \rho_{\rm drip}$, where $\rho_x$ is
given by eqn.~(\ref{eq-rhox}). \ Shear waves with $\epsilon\propto
\rho^{-0.8}$. \ 
{\bf (3)} \ $\rho_{\rm rel} < \rho < \rho_x$, where 
$\rho_{\rm rel}\equiv (B^2/4\pi c^2)= 9\times 10^7 \, B_{15}^2$ 
gm cm$^{-3}$.  Non-relativistic Alfv\'en waves with 
$\epsilon \propto\rho^{0.25}$.
\ {\bf (4)} \ $\rho <\rho_{\rm rel}$.  Relativistic Alfv\'en waves
with $\epsilon = \delta B/B =$ constant. 
Thus $\epsilon (\rho_x)$ as given by 
eqn.~(\ref{eq-strain}) is the {\it maximum} strain in the crust above
the neutron drip layer.\footnote{BBGM did a careful study of the coupling
of seismic and Alfv\'en waves, including wave reflection effects. 
However, their numerical results for $B= 10^{11}$ G are in the regime
in which zone {\bf (3)} effectively does not
exist (i.e., $\rho_x$ occurs at a depth that is less than one wavelength 
for Alfv\'en waves in the relevant frequency range).}   

Linear Alfv\'en waves in a neutron star magnetosphere are accompanied by
oscillating currents, $\delta {\bf j} = (c/ 4\pi) 
(\nabla \times \delta{\bf B}) 
- (1/4\pi)(\partial \, \delta {\bf E}/ \partial t)$. \ 
If the ambient plasma density is
so low that the drift velocities needed to sustain $\delta {\bf j}$
exceed $c$, then the wave is {\it charge-starved}.  This occurs at
plasma densities less than
\begin{equation}
\label{eq-cstarve}
n_* \simeq {\nu \, (\delta B) \over 2 e c} = 1\times10^{11} \, L_{41}^{1/2}
\, \nu_3 \ \hbox{cm}^{-3},
\end{equation}
where $\nu = 10^{3} \, \nu_3$ Hz is the wave 
frequency.  [Note that BBGM predict a frequency spectrum peaked
at $\nu \sim 10^3 \, E_{41}^{-1/3}$ Hz for bursts triggered by crust
fractures.  This comes from the condition that
$\nu^{-1}$ be comparable to the shear wave crossing-time of the fracture.]
BBGM have speculated that charge-starved Alfv\'en waves accelerate charges
until they are limited by radiation losses, thus producing GRBs as 
the waves undergo nonlinear damping. If this is true, then GRB emission
from ordinary radiopulsars might occur only after they cross the ``death
line" at age $\sim 2\times10^7 \, B_{12}^{-0.95}$ yrs (Chen \& Ruderman
1993), after which time the ambient plasma density in the magnetosphere   
is less than $n_*$. \ This could produce a delayed turn-on of GRBs
from pulsars, as in the galactic halo model of Li \& Dermer (1992).
However, the energy requirements for bursters (eq.~[\ref{eq-teng}] and 
eq.~[\ref{eq-breq}]),
the crust--crumbling limit (eq.~[\ref{eq-waveB}]) and the over-concentration
of old pulsars to the galactic disk are all potential difficulties
with this scenario (LD92).  Magnetars, which may avoid 
these drawbacks,  cross the death line at 
$t_D \sim 2 \times 10^5 \, B_{14}^{-1}$ yrs, comparable to the length
of the SGR activity epoch.    Subsequently, there 
exists a co-rotation charge density in the magnetosphere
of magnitude (Goldreich \& Julian 1969)
\begin{equation}
\label{eq-corotdens}
n_{\scriptscriptstyle GJ} \simeq {B\over P_{\rm rot} \, e \, c} \simeq
1\times 10^{12} \ B_{14}^{1/2} \ \left({t\over t_D}\right)^{-1/2} \
\hbox{cm}^{-3},
\end{equation}
where we have assumed
a value of $P_{\rm rot}(t)$ appropriate for spindown driven
by a constant dipole field.  Comparison with eq.~(\ref{eq-cstarve}) shows 
that, just after crossing the death line, Alfv\'en waves carrying
luminosities characteristic of halo GRBs are {\it not} charge-starved
near a magnetar's surface. The waves become charge-starved only after 
a time period
\begin{equation}
\label{eq-magstarve}
t_* = 4 \times 10^7 \ L_{41}^{-1} \ \nu_3^{-2} \ \hbox{yrs},
\end{equation}
independent of $B$ in this high--$B$ domain.  Interestingly, this is close
to the turn-on time required by LD92. \  The connection of charge-starved
Alfv\'en waves to GRBs is uncertain, and merits further study.

\clearpage

\centerline{FIGURE CAPTIONS}
\bigskip
\begin{figure}[htbp]
\caption{ Schematic diagram of a beamed gamma-ray burster 
escaping from the Galaxy. The Galaxy is depicted in cross-section
by a bulged disk, oriented horizontally at the bottom of the figure. 
The burster is born at point {\bf A} in the galactic disk, with
a recoil velocity ${\bf V_r}$ in a random direction. 
Gamma ray emissions occur within indicated 
cones of angular radius $\phi_b$ about the star's magnetic axis 
${\bf \mu}$, which we assume is nearly aligned with ${\bf V_r}$. \  
As the star sails into the halo, reaching point {\bf B} and beyond, 
the line of sight from Earth to the star 
gets increasingly aligned with ${\bf V_r}$. \
Thus more and more escaping bursters
become visible at larger distances $r$ from the galactic center. 
The {\it fraction} of all bursters which are detectable
increases with distance as the transverse area of the beaming cone 
in the galaxy, or $\sim r^2$. \ This tends to counter-balance the
free-streaming density trend,  $n \propto r^{-2}$  within
a ``core radius" $R_c \sim R_o/\phi_b$, where $R_o\sim 10$ 
kpc is a typical galactic disk dimension. 
At distances larger than $R_c$, all bursters are 
detectable at Earth, and the $n\propto r^{-2}$ 
density trend prevails. 
Because only a tiny fraction $1- \hbox{cos}\phi_b \simeq 0.06 \, 
(\phi_b/ 20^{\scriptscriptstyle \circ})^2$ of all bursters are
visible when they are born in the galactic disk, the observable
dipole anisotropy of GRB positions, 
$\cost$, is greatly reduced compared 
to what it would be if bursts were emitted isotropically.
\label{fig-hbm-concept}}
\end{figure}

\begin{figure}[htbp]
\caption{The 2-D cross section of the ``zone of avoidance'' (ZOA)
for beaming angles $\phi_b = 10^{\circ}, \, 20^{\circ}$
and $40^{\circ}$. The thick horizontal line represents the
galactic plane where the galactic center (GC) and Earth are located.
The circles are loci {\it within} which bursting stars born at GC
and moving in straight lines are {\it not} detectable at Earth. 
Everywhere within these ZOA circles, the angles between the 
star--to--Earth and star--to--GC directions are larger than $\phi_b$.
A star becomes detectable either when it travels beyond the circle, or 
when it is inside the lens-shaped overlap region of circles 
directly between GC and Earth.
The radius of the circles are $D_{\rm sun} / \hbox{sin} \phi_b$,
where $D_{\rm sun}=8.5$ kpc.
\label{fig-zoa}}
\end{figure}

\begin{figure}[htbp]
\caption{ Parmeters describing the 
spatial distribution of detectable bursters in the HBM are plotted 
as a function of distance from Earth, for several different beaming angles
$\phi_b$: $90^\circ$, or unbeamed {\it (thick solid line)};
$40^\circ$ {\it (thin solid line)}; $20^\circ$ {\it (dotted)}; 
$10^\circ$ {\it (dashed)} 
and $5^\circ$ {\it (alternating short and long-dashed).}  The 
{\it thick long-dashed}
horizontal lines in each subplot are the expected values if the distribution
is isotropic at all distances. Filled circles, placed
at $D=180$ kpc in each subplot, are current 
BATSE values for 1122 bursts; the error bars show 
3--$\sigma$ statistical uncertainties. Models with 
$\phi_b\geq 40^\circ$ or $\phi_b\leq 10^\circ$ are evidently 
ruled out by current observational limits, 
but a significant range of BATSE sampling depths $D_B \gsim 100$ kpc
are allowed if $\phi_b \sim 20^\circ$.  
\label{fig-ang-hbm} }
\end{figure}

\begin{figure}[htbp]
\caption{Comparison of the intensity distribution of the HBM
with combined PVO and BATSE data. Note the homogeneous, Euclidean ``core''
with slope $-3/2$ at the bright end, which turns over to a
significantly milder slope at the faint end.  A beaming angle 
$\phi_b=20^{\circ}$ is used in the HBM plotted here, and
the (standard-candle) intrinsic burst luminosity is 
varied to make the fitting, which has
a $\chi^2 \sim 6.9$ with 12 degrees of freedom. The implied peak
luminosity, including the beaming reduction factor, is 
$\sim 1.5\times 10^{40}$ ergs s$^{-1}$.
\label{fig-lognp-pbh}}
\end{figure}

\begin{figure}[htbp]
\caption{ Comparison of the intensity distribution of HBM
with BATSE only. Again, $\phi_b = 20^{\circ}$ but the derived peak
luminosity from this fitting is slightly smaller than that in 
Figure \ref{fig-lognp-pbh}.
\label{fig-lognp-bh}}
\end{figure}

\begin{figure}[htbp]
\caption{Cumulative plots of the angular $($an$)$isotropy measures 
versus peak photon flux $P_{256}$, for bursts in the 3B catalog 
(814 in total).  The solid curve is the HBM prediction
for $V_r = 1000$ km s$^{-1}$ and $\phi_b = 20^\circ$;
data points are from BATSE; the dashed line shows expected 
values if the bursts are distributed isotropically on the sky.  
The intrinsic burst peak luminosity found by fitting the 
log ${\cal N}$---log $P$ distribution was used to fix the 
the HBM curves' horizontal positions.
Error bars on the BATSE points show 1--$\sigma$ 
statistical uncertainties appropriate for comparing the data with the HBM. \ 
Both models (HBM and isotropy) are 
corrected for BATSE's imperfect sky coverage. 
Isotropy gives a better fit to the data than does the HBM. \
However, the data are consistent at the $\lsim 2$--$\sigma$ level or better 
with {\it both} models, at all observed flux levels. 
\label{fig-ang-hbm-3b}}
\end{figure}

\begin{figure}[htbp]
\caption{Gamma-ray intensity, $I(\theta)$, is plotted as a function of angle 
$\theta$ away from the beam axis, for the relativistic beam model of 
Mao \& Yi (1994).
In this model, luminous material is expanding uniformly into a cone of 
half opening-angle $\Delta \theta$ with Lorentz factor $\gamma$, and emitting 
photons isotropically in the material rest frame. 
In this plot we keep $\Delta\theta = 20^{\circ}$, while varying $\gamma$
as shown in the figure legend. 
Also shown is the $\phi_b = 20^\circ$ uniform beam case {\it (solid line),} 
which is equivalent to $\Delta\theta = 20^\circ$ and $\gamma = \infty$.
\label{fig-beam-int}}
\end{figure}

\begin{figure}[htbp]
\caption{The amounts by which observable GRB angular statistics in 
galactic halo models with variable beaming patterns $I(\theta)$ 
{\it differ} from their values in the  uniform beam (``unibeam") HBM, 
$\phi_b = 20^\circ$, are 
plotted as a function of peak photon flux. \ 
The beam intensity models considered here have  
$\Delta \theta = 20^\circ$ and $\gamma=3, 5, 10$ as indicated in the
figure legend; corresponding 
$I(\theta)$ functions are shown in Figure 7.  
The curves are not completely smooth because of the finite number of
stars used in the Monte Carlo simulations.
\label{fig-ang-diff}}
\end{figure}

\begin{figure}[htbp]
\caption{Similar to Figure \ref{fig-ang-hbm-3b} but including
the effects of luminosity functions.
Filled dots are from BATSE;  3--$\sigma$ error bars are shown
at the faint end. The thick solid line is for the HBM assuming 
mono-luminosity GRBs (standard candles). The broken lines are HBM
results for two alternative luminosity functions discussed in the text:
$\phi (L) \propto L^{-3}$ for $L_{\rm min} < L < L_{\rm max}$ with
a dynamic range $L_{\rm max}/L_{\rm min} =20$ [{\it dotted line}]; and  
$\phi (L) \propto L^{3}$ 
with $L_{\rm max}/L_{\rm min} =4$ [{\it dashed line}].  
It is clear that a GRB luminosity function with a positive (negative) slope 
tends to improve (worsen) the HBM fit to BATSE angular data.
\label{fig-ang-hbm-lum-3b}}
\end{figure}

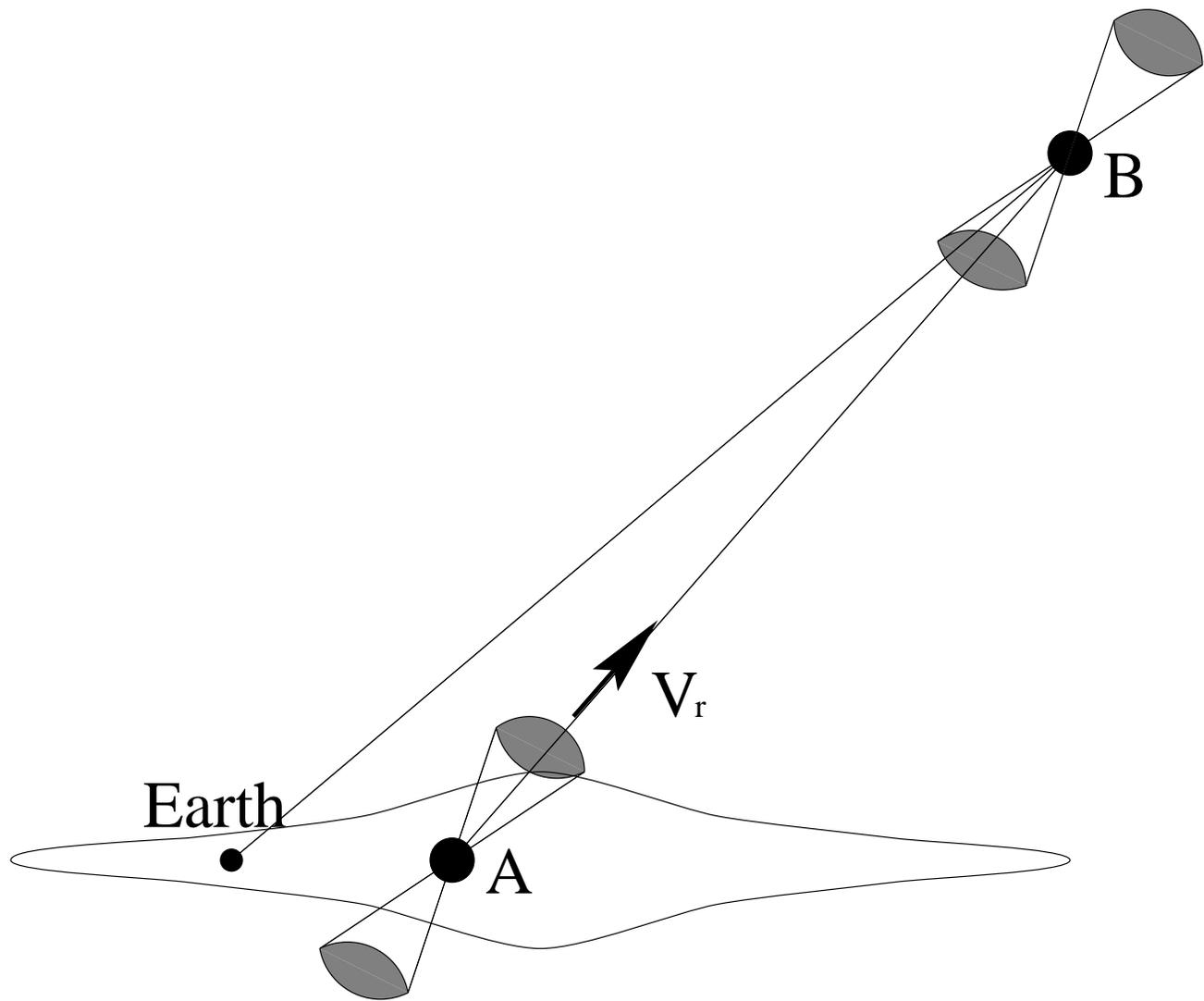
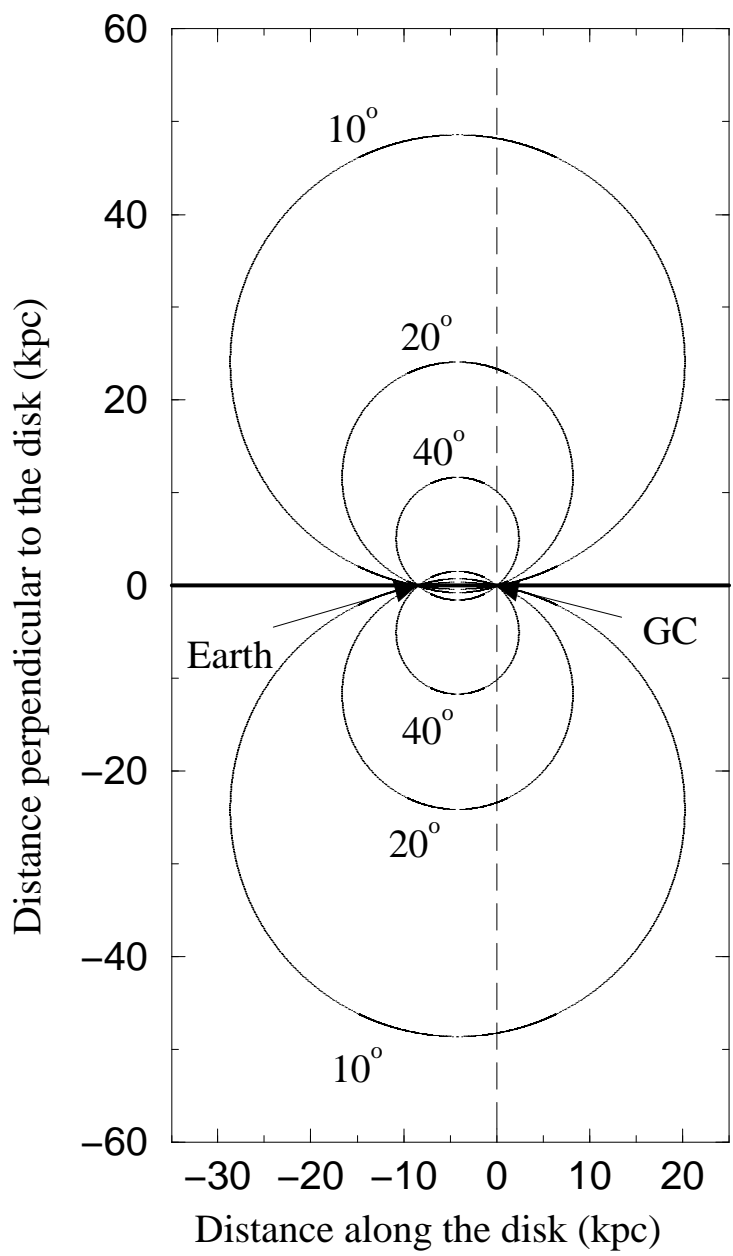
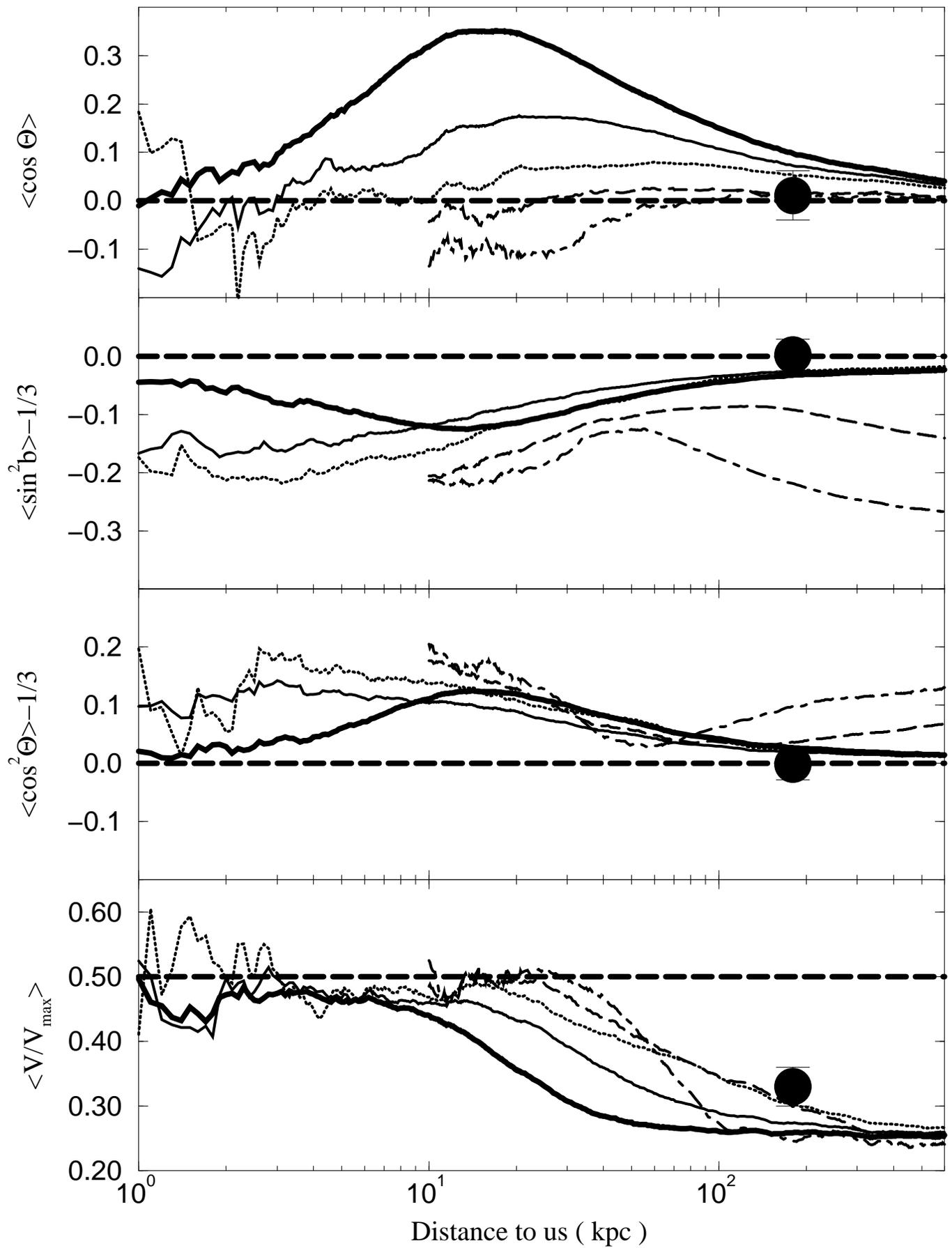
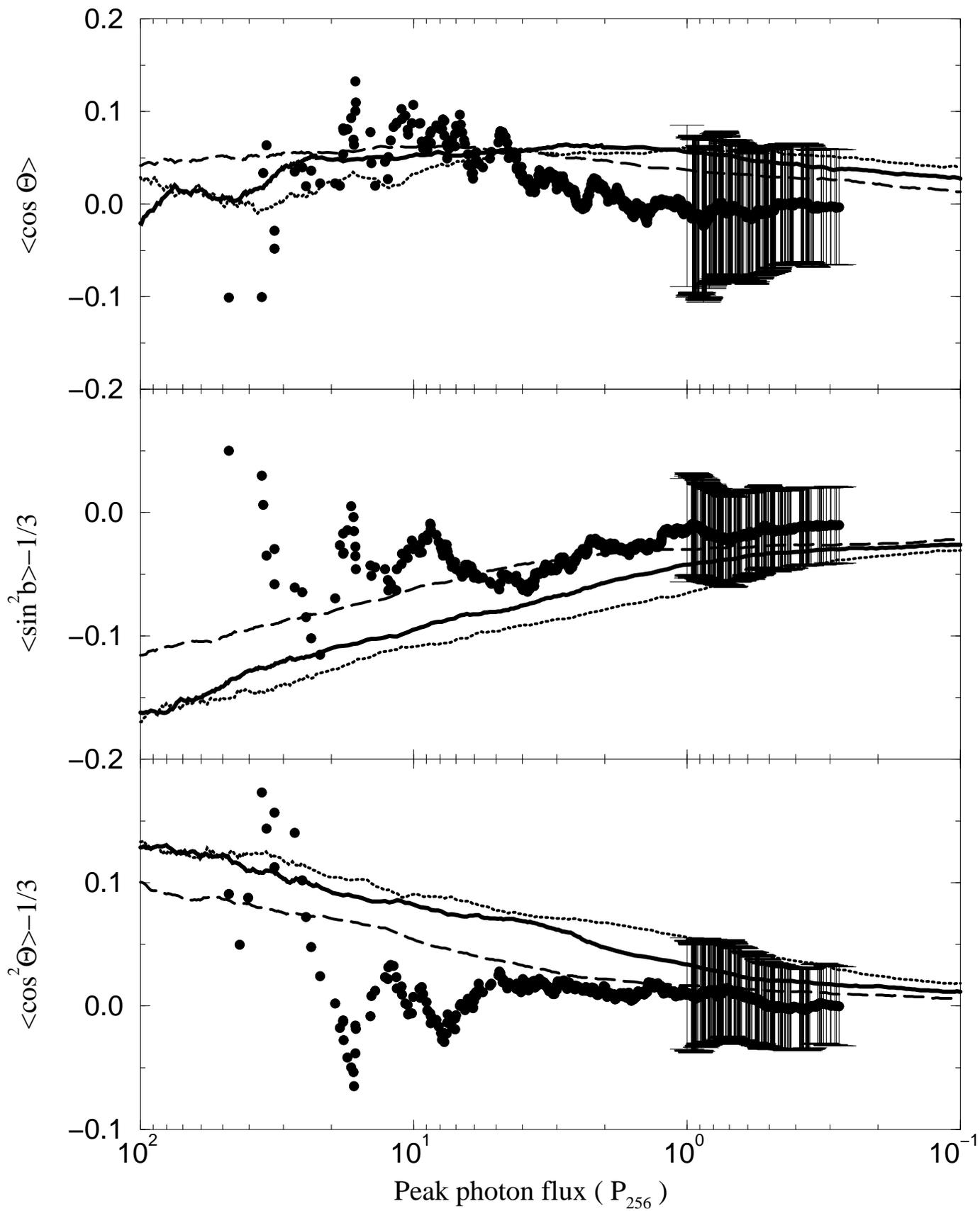
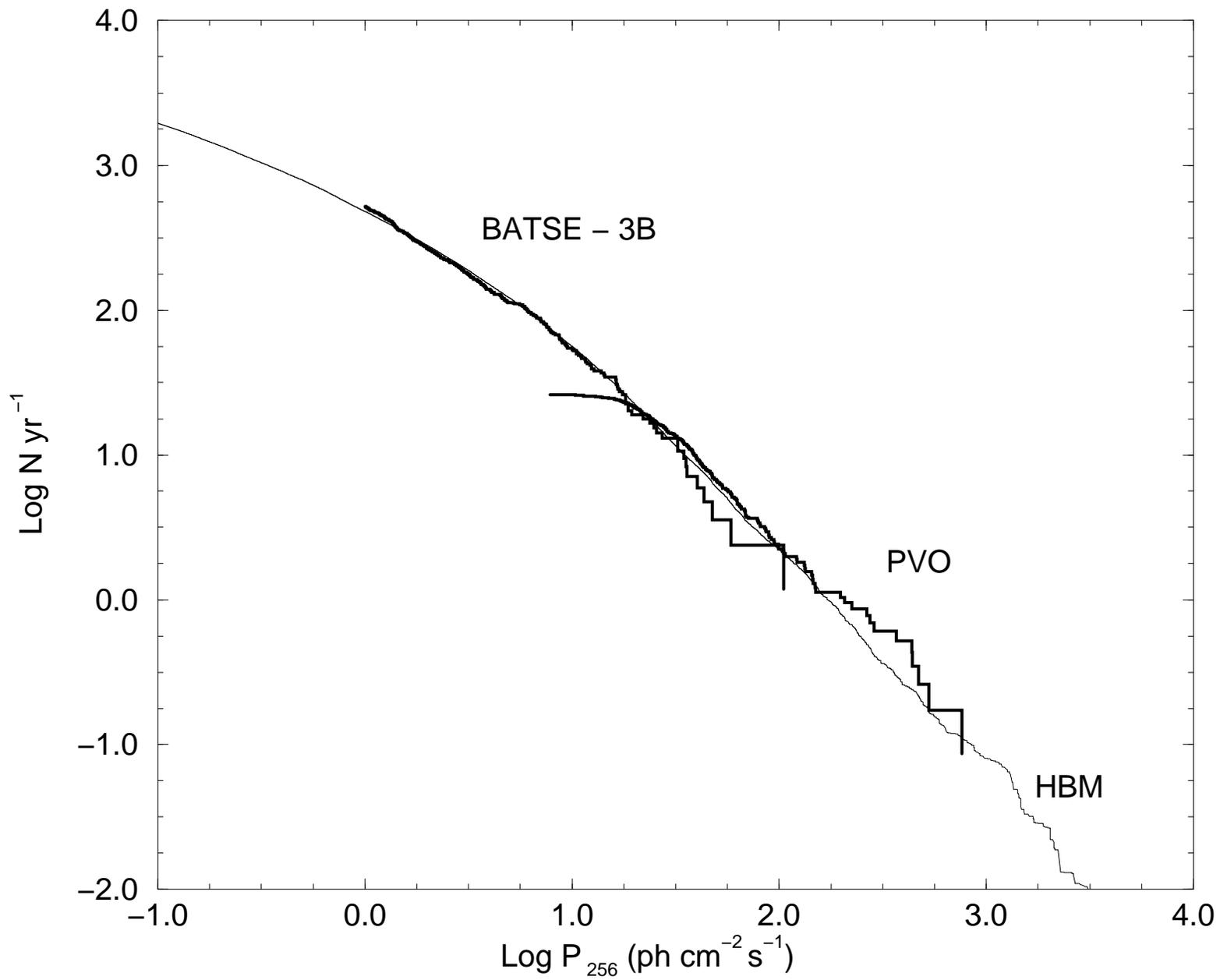
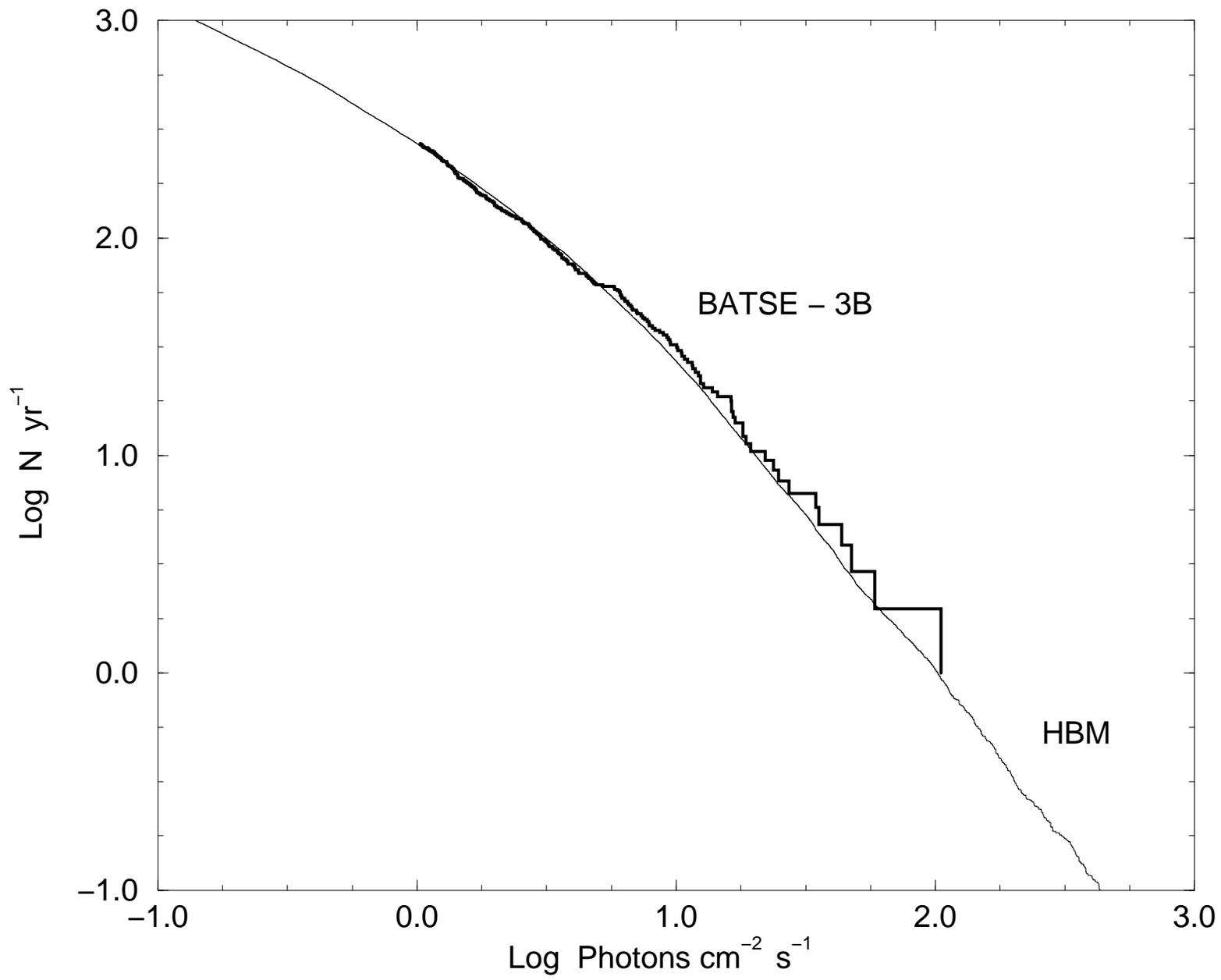
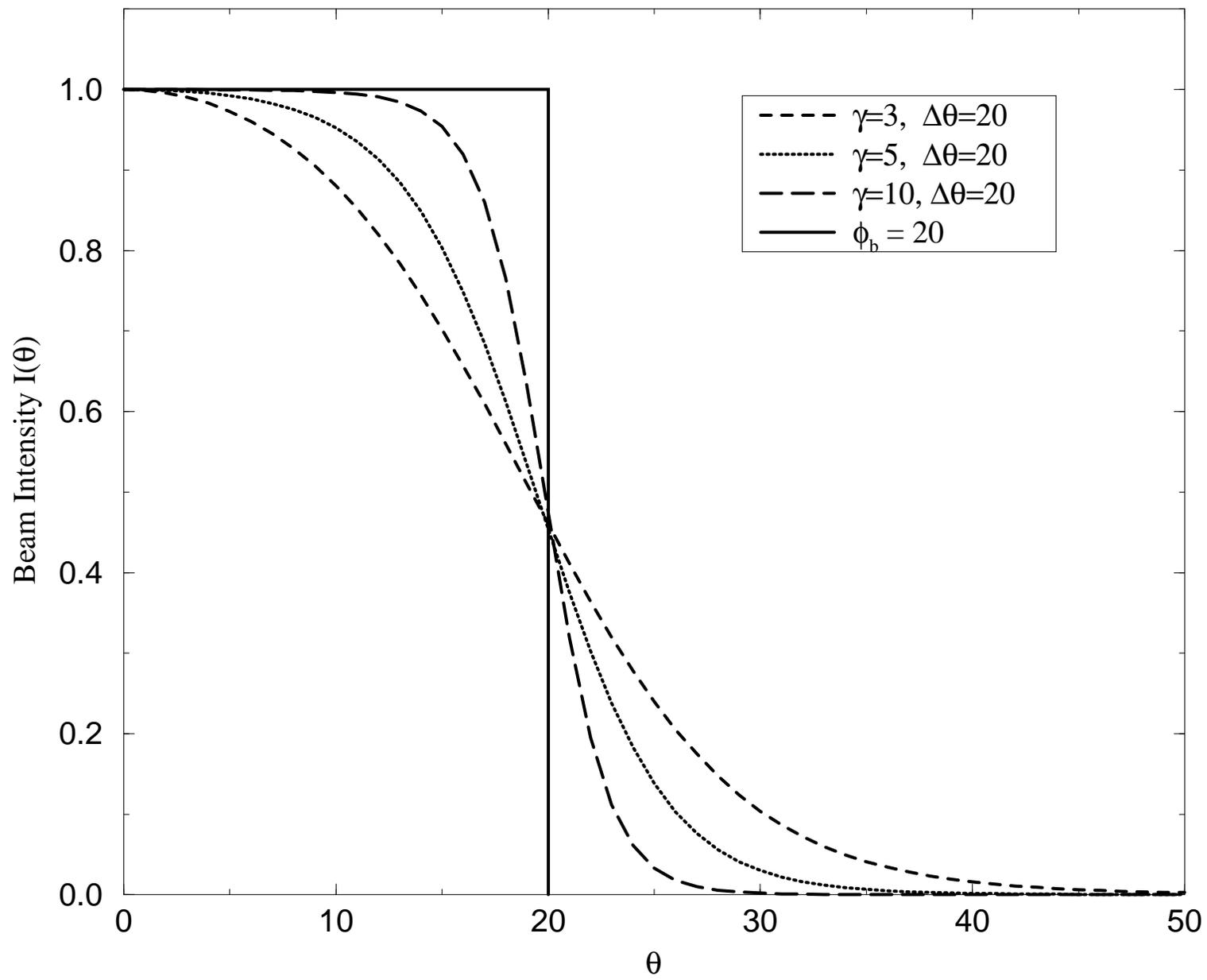
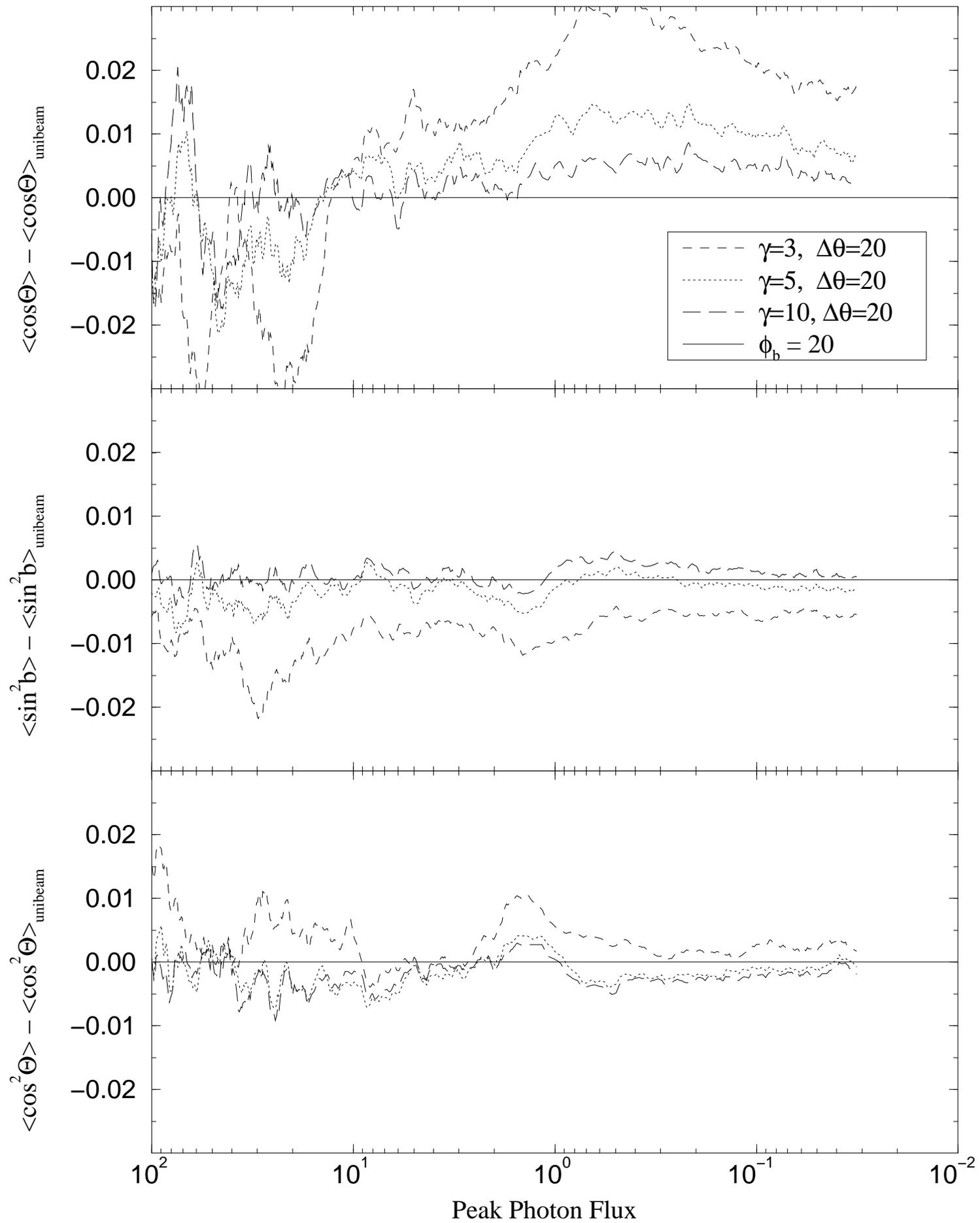
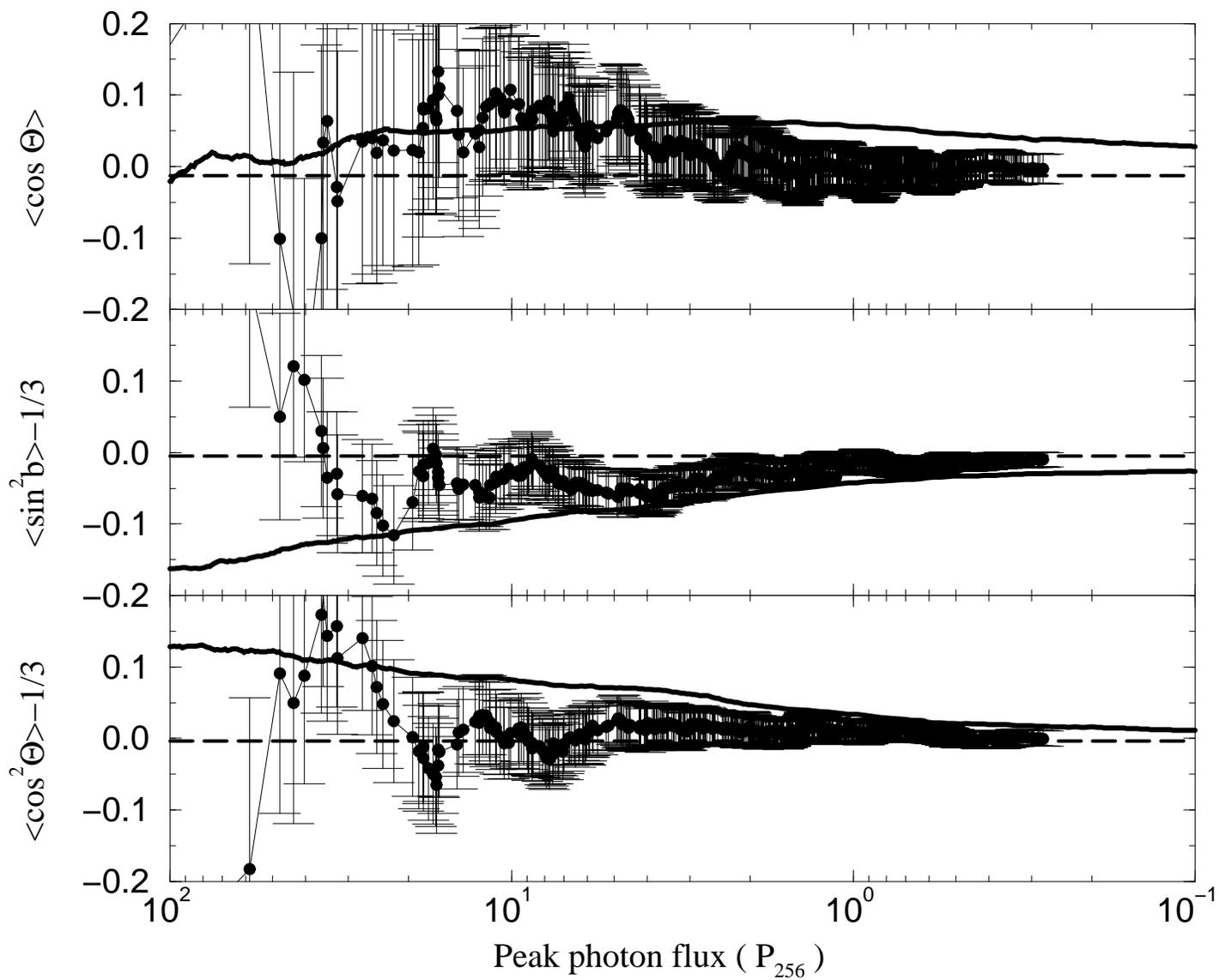
\end{document}